\documentclass[fleqn,twoside]{article} 		

\def\Title{	Toric Geometry and Calabi-Yau Compactifications}  
\def\Abstract{%
	These notes contain a brief introduction to the construction of 
	toric Calabi--Yau hypersurfaces and complete intersections with a
	focus on issues relevant for string duality calculations. The last
	two sections can be read independently and report on recent results 
	and work in progress, including torsion in cohomology,
	classification issues and topological transitions
}

\usepackage{epsf,multicol}	\usepackage{epsf}

\usepackage{cite}		\def\nvs{\vspace{-1pt}}

\def\ifundefined#1{\expandafter\ifx\csname#1\endcsname\relax}

\ifundefined{ENGLISHversion}	
				 \usepackage{ujp} 

\else	

					\makeatletter
\newcommand{\anot}[1]{{\noindent\footnotesize#1\\\rule{85.6mm}
{0.3pt}\vspace{4mm}\par}
             \noindent}

\def\udk#1{\def\@udk{#1}}
\def\@insti{}
\def\@adri{}
\def\@instii{}
\def\@adrii{}
\def\@instiii{}
\def\@adriii{}
\def\@instiiii{}
\def\@adriiii{}
\def\@instiiiii{}
\def\@adriiiii{}
\def\pusto{}
\def\nazva#1{\def\@nazva{#1}}
\def\nazvacol#1{\def\@nazvacol{#1}}
\def\avtor#1{\def\@avtor{#1}}
\def\avtorcol#1{\def\@avtorcol{#1}}
\def\inst#1{\def\@inst{#1}}\def\adr#1{\def\@adr{#1}}
\def\insti#1{\def\@insti{#1}}\def\adri#1{\def\@adri{#1}}
\def\instii#1{\def\@instii{#1}}\def\adrii#1{\def\@adrii{#1}}
\def\instiii#1{\def\@instiii{#1}}\def\adriii#1{\def\@adriii{#1}}
\def\instiiii#1{\def\@instiiii{#1}}\def\adriiii#1{\def\@adriiii{#1}}
\def\instiiiii#1{\def\@instiiiii{#1}}\def\adriiiii#1{\def\@adriiiii{#1}}
\newcommand {\maketitl}{%
\vbox{\hbox{\rule[1em]{36mm}{1pt}\hspace{4.6mm}\parbox[t]{135mm}%
{\raggedright\large\bf\@nazva\vspace{10pt}\\
\small\bf\@avtor}}
\vspace{10pt}\hbox{\raisebox{2pt}{\parbox[t]{40.6mm}
{\tiny\bf ~~~\vspace{-4pt}\\%
\raisebox{5pt}{\phantom{\bf\CYRU\CYRD\CYRK~\@udk}		}
\\ {\phantom{\bf\copyright 2001 \cyrr.}}}			}
\parbox[t]{135mm}%
{\bf\footnotesize\@inst\\\it\@adr%
\if \pusto\@insti\else\vspace{4pt}\\%
\bf\footnotesize\hbox to 0pt{\hss$\footnotesize^1$}\@insti\\%
\it\@adri\fi%
\if \pusto\@instii\else\vspace{4pt}\\%
\bf\footnotesize\hbox to 0pt{\hss$\footnotesize^2$}\@instii\\%
\it\@adrii\fi%
\if \pusto\@instiii\else\vspace{4pt}\\%
\bf\footnotesize\hbox to 0pt{\hss$\footnotesize^3$}\@instiii\\%
\it\@adriii\fi%
\if \pusto\@instiiii\else\vspace{4pt}\\%
\bf\footnotesize\hbox to 0pt{\hss$\footnotesize^4$}\@instiiii\\%
\it\@adriiii\fi%
\if \pusto\@instiiiii\else\vspace{4pt}\\%
\bf\footnotesize\hbox to 0pt{\hss$\footnotesize^5$}\@instiiiii\\%
\it\@adriiiii\fi%
\vspace{4mm}}%
}
\hbox{\rule{\textwidth}{1pt}}\vspace{4mm}
}
}
\renewcommand{\@evenhead}%
{\raisebox{0pt}[\headheight][0pt]%
{\vbox{\hbox to\textwidth{%
\tiny\MakeUppercase\@avtorcol\strut\hfil}\hrule height 0.76pt}}%
}
\renewcommand{\@oddhead}%
{\raisebox{0pt}[\headheight][0pt]%
{\vbox{\hbox to\textwidth{%
\hfil\strut\tiny\MakeUppercase\@nazvacol}\hrule height 0.76pt}}%
}
\renewcommand{\@evenfoot}{\small{\large\bf\thepage} \hfil\it 	\phantom{ISSN%
~0503-1265. \CYRU\cyrk\cyrr. \CYRF\cyrii\cyrz.
\CYRZH\cyru\cyrr\cyrn. 2001. \CYRT. 46, N 12}			}
\renewcommand{\@oddfoot}{{\small\it \phantom{ISSN~0503-1265.
\CYRU\cyrk\cyrr. \CYRF\cyrii\cyrz. \CYRZH\cyru\cyrr\cyrn. 2001. \CYRT. 46, N 12}}%
\hfil \large\bf\thepage}

\def\one{%
\end{multicols}
\par\noindent\raisebox{1.5em}{%
\rule{0.5\textwidth}{0.3pt}}\hspace{0.5\textwidth}\par}
\def\two{%
\par\noindent\hspace{0.5\textwidth}\raisebox{0em}{%
\rule{0.5\textwidth}{0.3pt}}
\begin{multicols}{2}
\noindent}
\renewcommand\section{\@startsection {section}{1}{0pt}%
                                    {-3.5ex \@plus -1ex \@minus -.2ex}%
                                    {2.3ex \@plus.2ex}%
                             {\raggedright\normalfont\normalsize\bfseries}}

 \renewcommand\subsection{\@startsection{subsection}{2}{\z@}%
                                      {-3.25ex\@plus -1ex \@minus -.2ex}%
                                      {1.5ex \@plus .2ex}%
                                      {\normalfont\normalsize\bfseries}}
 \renewcommand\subsubsection{\@startsection{subsubsection}{3}{\z@}%
                                      {-3.25ex\@plus -1ex \@minus -.2ex}%
                                      {1.5ex \@plus .2ex}%
                                      {\normalfont\normalsize\bfseries}}

\renewcommand\paragraph{\@startsection{paragraph}{4}{\z@}%
                                     {3.25ex \@plus1ex \@minus.2ex}%
                                     {-1em}%
                                     {\normalfont\normalsize\bfseries}}
 \renewcommand\subparagraph{\@startsection{subparagraph}{5}{\parindent}%
                                        {3.25ex \@plus1ex \@minus .2ex}%
                                        {-1em}%
                                       {\normalfont\normalsize\bfseries}}
 \renewcommand \thesection {\@arabic\c@section{.}}
 \renewcommand\thesubsection   {\thesection\@arabic\c@subsection}
 \renewcommand{\@biblabel}[1]{#1.\hfill}
 \renewenvironment{thebibliography}[1]
    {\vspace{12pt plus 2pt minus 2pt}\footnotesize%
     \list{\@biblabel{\@arabic\c@enumiv}}%
          {\settowidth\labelwidth{\@biblabel{#1}}%
            \leftmargin\labelwidth
            \advance\leftmargin\labelsep
            \@openbib@code
            \usecounter{enumiv}%
            \let\p@enumi\@empty
            \renewcommand\theenumiv{\@arabic\c@enumiv}}%
      \sloppy
      \clubpenalty4000
       \@clubpenalty \clubpenalty
       \widowpenalty4000%
       \sfcode`\.\@m}
      {\def\@noitemerr
        {\@latex@warning{Empty `thebibliography' environment}}%
       \endlist}
\makeatother

\fi	

\usepackage[cp866nav]{inputenc}			
\usepackage[english,ukrainian]{babel}
\usepackage{mathrsfs}

\usepackage{amssymb}
\usepackage[dvips]{graphicx}			
\usepackage{amsthm}
\usepackage{amsfonts, amscd}
\usepackage{amssymb}

\usepackage{stmaryrd}
\usepackage{eufrak}
\usepackage{bbold}			

\nazva{\Title			
}%

\udk{}

\nazvacol{\Title}	\avtor{Maximilian KREUZER}

\avtorcol{M. KREUZER}

\inst{Institute for Theoretical Physics}
\adr{Vienna University of Technology, Wiedner Hauptstr. 8-10, 1040 Vienna, 
	Austria 		\\e-mail: Maximilian.Kreuzer@tuwien.ac.at}

\let\a=\alpha	\let\b=\beta	\let\g=\gamma	\let\d=\delta 	
 	\let\h=\eta 	\let\th=\theta 	\let\e=\varepsilon
	\let\l=\lambda	\let\m=\mu	
\let\n=\nu	\let\x=\xi 	\let\p=\pi 	\let\r=\rho	\let\s=\sigma 
\let\t=\tau 	\let\o=\omega 	\let\c=\chi	\let\ph=\varphi \let\ps=\psi 
\let\Ph=\phi 	 	 	\let\O=\Omega 	\let\S=\Sigma 
	 	 	\let\G=\Gamma 	\let\D=\Delta

\let\0=\over      \let\1=\vec      \def\2{{1\over2}}    \let\3=\ss
\let\4=\underline \let\5=\overline \let\6=\partial      \def\7#1{{#1}\llap{/}}
\def\8#1{{\textstyle{#1}}}         \def\9#1{{\ifmmode{\pmb{#1}}\else\bf#1\fi}}

\def\IC{{\mathbb C}} \def\IN{{\mathbb N}} \def\IP{{\mathbb P}} 
\def\IR{{\mathbb R}} \def\IZ{{\mathbb Z}} \def\IQ{{\mathbb Q}}

\let\bra=\langle  \let\ket=\rangle	\let\hc=\dagger  \let\ex=\times
\def\mao#1{\mathop{\rm #1}\nolimits}	\let\minus=\setminus \def\minus{-}
\def\BE {\begin{equation}} \def\EE {\end{equation}} 
\def\BEA{\begin{eqnarray}}   \def\EEA{\end{eqnarray}} 
\def\BP{\begin{picture}} \def\EP{\end{picture}}		\let\then=\Rightarrow

		\def\eg{{\mathfrak g}}
\def\nb{{\bar\n}}		
	\long\def\del#1\enddel{}

\long\def\comment#1{{\BP(0,0)(0,0)\unitlength=1mm\put(-15,5.7){
        \makebox(0,0)[tl]{\parbox{187mm}{\tiny\baselineskip=18pt #1}}}\EP}}

\def\rel#1 #2{\buildrel #1 \over {#2}}  \def\nibf#1:{\noindent{\bf #1:}}

\usepackage{epic}
\def\putlin#1,#2,#3,#4,#5){\put#1,#2){\line(#3,#4){#5}}} 
\def\putlab#1)#2#3{\put#1){\makebox(0,0)[#2]{\small #3}}}

\def\makeYZ(#1,#2,#3){
	\Xaux=#1 \multiply \Xaux by \XYfac \Yaux=#2 \multiply \Yaux by \Yfac 
	\advance\Yaux by -\Xaux
	\Xaux=#1 \multiply \Xaux by \XZfac \Zaux=#3 \multiply \Zaux by \Zfac 
	\advance\Zaux by -\Xaux
}
\def\xyzline(#1,#2,#3)(#4,#5,#6){\makeYZ(#1,#2,#3) \YA=\Yaux \ZA=\Zaux 
	\makeYZ(#4,#5,#6)	\drawline(\YA,\ZA)(\Yaux,\Zaux)
}				

\def\subdef#1{\gdef\globalColor##1{##1}}	\def\CFigOffset{4}       
\def\TC#1{\subdef{cmyk #1}}      \subdef{Black}
\def\red{\TC{0 1 1 0}} \def\blue{\TC{1 1 0 0}} \def\black{\TC{0. 0. 0. 1.}}
\def\lila    {\TC{0. 1. 0. 0.}}			\def\HS#1 {\hspace*{#1pt}}
			
\def\plb#1 #2 {Phys. Lett. {\bf B#1} #2 }	
\def\npb#1 #2 {Nucl. Phys. {\bf B#1} #2 }
\def\jgp#1 #2 {J. Geom. Phys. {\bf #1} #2 }
\def\atmp#1 {Adv. Theor. Math. Phys. {\bf #1} }
\def\jhep#1 {J. High Energy Phys. {\bf #1} }

\ifundefined{draftmode}	\else	\input draftmode.tex
   \def\LLab#1{\BP(0,0)\unitlength=1mm\put(-3,-1.6){\makebox(0,0)[cr]{\tiny #1
        \rlap{$_{_{\makeatletter\csname TRef#1\endcsname\makeatother}}$}}}\EP}
	\fi

\begin{document}           

\maketitl                 

\begin{multicols}{2}

\anot{\Abstract}


\section*{Introduction}

Toric geometry is a beautiful part of mathematics that relates discrete 
and algebraic geometry and provides an elegant and intuitive construction 
of many non-trivial examples of complex manifolds 
\cite{Dan,Oda,Ful,Ewald,Cox,CoxMini}. Sparked  
by Batyrev's construction of toric Calabi--Yau hypersurfaces, 
which relates mirror symmetry to a combinatorial duality of convex polytopes
\cite{Bat}, toric geometry also became a pivotal tool in string theory. 
It provides efficient tools for the construction and analysis of
large classes of models, and for
computing quantum cohomology and symplectic invariants 
\cite{Cox:vi,HKPTVZ,KKRS,TopRev}, 
fibration structures \cite{AKMS,fft,Hu:2000,Rohsie}
for non-perturbative dualities \cite{KachruV,KLM,AL,CanFont},
and lagrangian submanifolds for open string and D-brane physics 
\cite{SYZ,AgaVafa,Coni}. $F$-theory compactifications \cite{F,F12}, which 
are based on elliptic Calabi--Yau 4-folds, is maybe the most 
promising approach to realistic unified string models for particle physics
\cite{Denef}, but also 3-folds with torsion in cohomology have been used
for successful model building \cite{Z3xZ3}. 

In the present notes we describe some tools that are provided by
toric geometry and report on some recent results. 
Section \ref {Sec:TG} contains the basic definitions and constructions
of toric varieties, working mainly with the homogeneous coordinate ring. 
In section \ref {Sec:CY} we recall the string theory 
	context in which Calabi--Yau 
	geometry becomes important for particle physics and describe
the	toric construction of hypersurfaces
	and complete inter\-sections. 
Section \ref {Sec:FibTor} explains fibrations 
	and torsion in cohomology 
	in terms of the combinatorics of polytopes.
In section \ref{Sec:Conc} we summarize recent results and work in progress
and conclude with a list of open problems.


\section{Basics of Toric Geometry}	\label{Sec:TG}

The (algebraic) $n$-torus is a product $T=(\IC^*)^n$ of $n$ co\-pies of the
punctured complex plane $\IC^*=\IC\setminus\{0\}$, which we regard as a
multiplicative abelian group. It is hence a complexification of the real
$n$-torus $U(1)^n$.
A toric variety is defined as a (partial) compactification of $T$ in 
the following sense:
It is a (normal) variety $X$ that contains an $n$-torus $T$ as a
dense 
open subset such that the natural action of the torus $T$ on itself extends to
an action of $T$ on the variety $X$.

The beauty of the toric geometry comes from the fact that the data
is encoded in combinatorial terms and that this structure 
can be used to derive simple formulas for sophisticated topological and 
geometrical objects. More precisely, the data is given by a fan $\S$, which 
is a finite collection of strongly convex rational polyhedral cones (i.e. 
cones generated by a finite number of lattice \hbox{points} and not 
containing a complete line) such that all faces of cones and all 
intersections of any two cones also belong to the fan. 

The space in which $\S$ lives can be obtained as follows \cite{CoxMini}: 
If we parametrize the torus by coordinates $(t_1,\ldots,t_n)$
the character group $M=\{\c:T\to\IC^*\}$  of $T$ can be identified with a
lattice $M\cong \IZ^n$, where $m\in M$ corresponds to the character
$\c^m((t_1,\ldots,t_n))=t_1^{m_1}\ldots t_n^{m_n}\equiv t^{m}$.
Another natural lattice that comes with the torus $T$ can be identified
with the group of algebraic one-parameter subgroups,
$N\cong\{\l:\IC^*\to T\}$ where
$u\in N$ corresponds to the group homomorphism
$\l^u(\t)=(\t^{u_1},\ldots,\t^{u_n})\in T$ for $\t\in\IC^*$.
The composition $(\c\circ\l)(\t)=
\c(\l(\t))=\t^{\bra\c,\l\ket}$ 
defines a canonical
pairing $\bra\c^m,\l^u\ket=m\cdot u$ which makes $N$ and 
$M\cong\mao{Hom}(N,\IZ)$ a dual pair of lattices 
(or free abelian groups). The
characters $\c^m$ for $m\in M$ can be regarded as holomorphic functions on
the torus $T$ and hence as rational functions on the toric variety $X$.
We will see that the lattice $N$, or rather its real extension 
$N_\IR=N\otimes_\IZ\IR\cong\IR^n$, is where the cones $\s\in\S$ of the 
fan live.

We first construct the rays $\r_j$ with $j=1,\ldots,r$, i.e. the 
one-dimensional cones $\r_j\in\S^{(1)}$ of the fan where
the $n$-skeleton $\S^{(n)}$ 
contains the $n$-dimensional cones of $\S$.
Recall that 
divisors are formal linear combinat\-ions of subvar\-iet\-ies of $X$ of 
(complex) codimension 1, i.e. of dimension $n-1$.
It can be shown that normality of $X$ implies that the divisors 
$\mao{div}(\c^m)$, i.e. the hypersurfaces defined by the equations $\c^m=0$, 
are equal to sums $\sum_1^r a_jD_j$ for some finite set of {irreducible 
divisors $D_j$}. Like $\mao{div}(\c^m)$ the $D_j$ are $T$-invariant
and hence unions of complete orbits of the torus action. 
The coefficients $a_j(m)$ 
are unique so that the decomposition $\mao{div} \c^m\!=\!\sum a_jD_j$
defines linear maps $m\!\to\! a_j(m)\!=\!\bra m,v_j\ket$. The irreducible
divisors $D_j$ hence define vectors $v_j\in N$ for $j\le r$ with 
$a_j(m)=\bra m,v_j\ket$. These vectors are the primitive generators of
the rays $\r_j$ that constitute the 1-skeleton $\S^{(1)}$ of the fan $\S$.
If we {locally write the equation of the divisor as $D_j=\{z_j=0\}$} 
with $z_j$ a section of some local line bundle then we can 
{write the torus coordinates, with appropriate choice of normalizations, 
as $t_i=\prod z_j^{\bra e_i,v_j\ket}$} on some dense subset of $T\subseteq X$.

\vspace{-5pt}
\subsection{Homogeneous coordinates} 		\label{HomoCoord}
\vspace{-2pt}
We now regard $\{ z_j\}$ as {global homogeneous coordinates}
$(z_1\!:\!\ldots\!:\!z_r)$ in generalization of the construction of 
the projective space
$\IP^n$ as a $\IC^*$-quotient of $\IC^{n+1}\setminus\{0\}$ via the
identification $(z_0\!:\!\ldots\!:\!z_n)\equiv(\l z_0\!:\!\ldots\!:\!\l z_n)$.
If all $z_j$ are non-zero 
then the coordinates
\BE						
	(\l^{q_1}z_1:\ldots:\l^{q_r}z_r)\sim (z_1:\ldots:z_r) \label{scaling}
	,\HS9 
	\HS9 \l\in\IC^*				
\EE
describe the same point of the torus $T$ 
with coordinates 				
\BE							
	t_i=\prod z_j^{\bra e_i,v_j\ket}  ~~~\in~~~T=X\setminus\bigcup D_j
\EE	
if $\sum  {q_j v_j=0}$, where
$v_j
$ are the generators of the rays $\r_j\in\S^{(1)}$ of the fan $\S$.
Since the vectors $v_j\in N$ belong to a lattice of dimension $n$
the scaling exponents $\{q_j\}\in \IZ^{\S^{(1)}}\cong\IZ^r$ in the 
identification (\ref{scaling})
are restricted by $n$ independent 
linear equations. Naively we 
therefore might expect that the toric variety $X$ can be written as  
a quotient
$	(\IC^r\setminus Z) \;/\;(\IC^*)^{r-n},
$
where $\{(z_j)\}=\IC^r\setminus Z$ is the set of allowed values for the 
homogeneous coordinates and the $(\IC^*)^{r-n}$--action on $\{(z_j)\}$ 
implements the identification (\ref{scaling}) with $\sum{q_j v_j=0}$.
The identification group corresponds, however,
to the kernel of the map $(z_j)\to (t_i=\prod z_j^{\bra e_i,v_j\ket})$ 
for $(z_j)\in(\IC^*)^r$.
If the $v_j$ do not span the $N$--lattice, i.e. if the quotient
\BE						
	G\cong N/(\mao{span}_\IZ\{v_1,\ldots,v_r\})
\EE
is a finite abelian group of order $|G|>1$, then this kernel 
is $(\IC^*)^r /(\IC^*)^n\cong(\IC^*)^{r-n}\ex G$ and contains
a discrete factor $G$.
The toric variety $X$ can hence be constructed in terms of 
homogeneous coordinates $z_j$, an exceptional set $Z\subset \IC^r$, and 
the identification group $(\IC^*)^{r-n}\ex G$ as		
\BE								\nvs
	X=(\IC^r-Z) \,/\,((\IC^*)^{r-n}\ex G),\label{QuotRep}	\nvs
\EE								
where the quotient can be shown to be ``geometrical'' if the fan $\S$ is 
simplicial \cite{CoxHCR} (cf., however, example 2 below).
\del
dealing with $\IZ$ modules rather than vector spaces, however, the quotient 
$\IZ^r/\IZ^n\cong \IZ^{r-n}\ex G$ may have a torsion part $G$. The torsion 
part $G$ is a finite abelian group that acts by phase transformations on the 
homogeneous coordinates. It is isomorphic to the lattice quotient 
i.e. to the quotient of $N$ by the sublattice generated by the $v_j$.
\enddel 
For a given set of generators $v_j\in\IR^n$ of the rays $\r_j\in\S^{(1)}$ 
we can construct different toric varieties by choosing different
lattices $N\subset \IR^n$ that contain the $v_j$ as primitive lattice vectors.
These are all abelian quotients of the variety for which
$N$ is the integral span of 
$\{v_j\}$.

The last piece of information that we need is the exceptional set $Z$. 
The limit points $(z_j)\in X\setminus T=\bigcup D_j$ that are added
to $T$ are determined by the conditions under \hbox{which} 
homogeneous coordinates are allowed to vanish. 
This is where the information of the fan $\S$ enters:  A subset of the 
coordinates $z_j$ is allowed to \hbox{vanish} simultaneously iff
there is a cone $\s\in\S$ containing all of the correspond\-ing rays $\r_j$.
In geometrical terms this means that the corresponding divisors $D_j$
intersect in $X$. The exceptional 
set $Z$ hence is the union of sets 		\nvs
\BE					
	Z_I=\{(z_1:\ldots:z_r)\,|\,z_j=0 \;\forall\; j\in I\} \label{PriCol} 
\EE						\nvs
for which there is no cone $\s\in\S$ such that $\r_j\subseteq\s$ for all
$j\in I$. 
Minimal index sets $I$ with this property are called {\it primitive
collections}. They correspond to the maximal irreducible components of 
$Z=\bigcup Z_I$. 

\begin{figure*}[htb]
\unitlength=1pt
\BP(260,60)(-52,-26)\unitlength=2pt
        \put(0,0){
\lila   \drawline(0,0)(-8,-8)\drawline(0,0)(-16,16)\drawline(-8,-8)(-16,16)
\black} \put(-25,-1){$\mathbb F_0$}\put(-25,7){$\mathbb F_1$}
        \put(-25,15){$\mathbb F_2$}
        \put(4,0){\matrixput(-18,-8)(8,0)3(0,8)4{\circle*2}
        \put(-2,0){\drawline(0,0)(-8,0)\drawline(-8,-8)(-8,8)
                \drawline(-16,16)(-8,0)}}
        \put(-8.7,-12.3){$z_0$}\put(4.5,-1){$z_1$}\put(-15,18){$z_2$}
        \put(-7,10){$w$}

        \put(10,9){$(\blue\l^2\red\m\black z_0\!:\!\blue\l\black z_1\!:\!
                        \blue\l\black z_2\!:\!\red\m\black w)$
                \normalsize$\in(\mathbb C^4\!-\!Z)/(\mathbb C^*)^2$}
        \put(20.5,0){$Z=\{z_0\!=\!w\!=\!0\}\cup\{z_1\!=\!z_2\!=\!0\}$}	\black
\EP
\def\ConeSigma{\drawline(0,10)(-20,40)\drawline(0,10)(20,40)
        \dashline2(0,10)(-5,50)\drawline(0,10)(5,30)
        \drawline(5,30)(20,40)\drawline(5,30)(-20,40)
        \drawline(-5,50)(20,40)\drawline(-5,50)(-20,40)}
\medskip\unitlength=1.2pt	
\BP(100,58)
\put(80,0){\ConeSigma \putlab(-5,52)b{$z_1$}\putlab(22,42)b{$z_3$}
	\putlab(-22,42)b{$z_2$}\putlab(4,32)b{$z_0$}	
	\putlab(0,4)t{\footnotesize(2b)}
	}	    
\put(20,0){\ConeSigma	\putlab(3,32)b{$\b$}\putlab(1,42)b{$\a$}
	\putlab(0,4)t{\footnotesize(2a)}
        \put(0,0){\blue\linethickness{2pt}\drawline(-20,40)(20,40)}\black}
\put(140,0){\ConeSigma	\putlab(9,47)b{$x$}\putlab(-14,46)b{$v$}
	\putlab(0,4)t{\footnotesize(2c)}
			\putlab(9,35)b{$u$}\putlab(-6,28)b{$y$}
        \put(0,0){\blue\linethickness{2pt}\drawline(-5,50)(5,30)}\black}
\EP

\stepcounter{figure}						\makeatletter
\immediate\write\@auxout{\string\newlabel{Hirze}{\thefigure}} 	\makeatother
Fig. \thefigure: The Hirzebruch surface $\mathbb F_2$ 
	as blowup  
 	of $W\IP_{211}$.
\HS4.5
\stepcounter{figure}						\makeatletter
\immediate\write\@auxout{\string\newlabel{ConiF}{\thefigure}} 	\makeatother
Fig. \thefigure: Toric desingularizations of the conifold.
\vspace*{-12pt}
\end{figure*}

\vspace{-5pt}
\subsection{Torus orbits}
\vspace{-2pt}
In terms of homogeneous coordinates the torus action amounts to the 
effective part of the $(\IC^*)^r$ action induced on the quotient 
(\ref{QuotRep}) by independent scalings of $z_j$ and thus extends from 
$T$ to $X$. The torus orbits into which $X$ decomposes are hence 
characterized by the index sets of the vanishing coordinates. It can be 
shown \cite{Ful} that the {\it torus orbits} $\cal O_\s$ are in one-to-one 
correspondence with the cones $\s\in\S$, where
${\cal O}_\s\cong (\IC^*)^{n-\dim\s}$ is the inter\-section of all divisors 
$D_j$ for $\r_j\subset\s$ with all complements of the remaining divisors.
The orbit $\cal O_\s$ is an open subvariety of the {\it orbit closure} 
$V_\s$, which is the intersection of all divisors $D_j$ for $\r_j\subset\s$
and which also has dimen\-sion $n-\dim \s$.

Further important sets are the {\it affine open sets} $U_\s$, 
\hbox{which} are the intersections of all complements $X\setminus D_j$ 
with $\r_j\not\in\s$ and
which provide a covering of $X$. The relations between these 
sets can be summarized as
\BE
	V_\s=\bigcup_{\t\supseteq\s}{\cal O}_\t,\HS15
	U_\s=\bigcup_{\t\subseteq\s}{\cal O}_\t,\HS15
	X=\bigcup_{\s\in\S} U_\s,
\EE
where $V_\s$ and $U_\s$ are disjoint unions and the last \hbox{union} can
be restricted to a covering of $X$ by the $U_\s$'s for maximal cones $\s$. 
In the traditional approach \cite{Dan,Oda,Ful,Ewald} a
toric variety $X$ is constructed by gluing its open
{\it affine patches} $U_\s$ 
along their intersections
\BE
	U_\s\cap U_\t=U_{\s\cap\t}\supseteq T=U_{\{0\}}.
\EE
The patches $U_\s$ are constructed in terms of their
rings of regular functions, which are generated by the characters $\c^m$
that are nonsingular on the relevant patch. Since 
\BE
	t_i=\prod z_j^{\bra e_i,v_j\ket}~~\then~~
	\c^m=t^m=\prod z_j^{\bra m,v_j\ket}		\label{Character}
\EE
the relevant exponent vectors $m\in M$ are the lattice points in the dual cone
\BE
	\s^\vee						      \label{DualCone}
	=\{x\in M_\IR\;:\;\bra x,v\ket\ge0~~\forall~v\in\s\}.
\EE
More abstractly, the semigroup algebra $A_\s=\IC[\s^\vee\cap M]$ of
the semigroup $\s^\vee\cap M$ is, by definition, the ring of regular 
functions on $U_\s$, so that the points of $U_\s$ 
can be obtained as the spectrum $\mao{Specm}(A_\s)$ 
of maximal ideals. The Zariski topology of $U_\s$ can be constructed in terms 
of the prime ideals 
and
the gluing can be worked out by relating the characters in
different patches (cf. example 2 below).
We now state two important theorems \cite{Dan,Oda,Ful,Ewald}:
\\[4pt]
\nibf	Theorem 1: A toric variety is {\it compact} 
	if and only if the fan is complete, i.e. if the 
	support of the fan covers the $N$ lattice $\;|\S|=\bigcup_{_{
	\S}}\! \s=N_\IR$.
\\[3pt]
We prove the {\it only if}: For an incomplete fan we consider some
$u\in N\setminus|\S|$ and a one-parameter family of points 
$p_\l=(\l^{u_1}t_1,\ldots,\l^{u_n}t_n)\in T$. Evaluation of 
$\c^m$ yields $\c^m(p_\l)=\l^{\bra m,u\ket}\c^m(p_1)$.
But the limit point $p_{\l\to0}$ cannot be contained in any patch 
$U_\s$ because $\c^m(p_\l)$ diverges as $\l\to0$ for $m\in\s^\vee$ 
and $u\not\in\s$ so that $\bra m,u\ket<0$.
\\[4pt]
\nibf 	Theorem 2: A toric variety is {\it non-singular} if and only if all 
	cones are simplicial and basic, i.e. if all 
	cones $\s\in\S$ are generated by a subset of a lattice basis of $N$. 
\\[3pt]
To illustrate these theorems we work out two examples:\vspace{3pt}

\del
\BP(80,60)(-37,-8)\unitlength=2pt
        \put(0,0){
\lila   \drawline(0,0)(-8,-8)\drawline(0,0)(-16,16)\drawline(-8,-8)(-16,16)
\black} \put(-25,-1){$\mathbb F_0$}\put(-25,7){$\mathbb F_1$}
        \put(-25,15){$\mathbb F_2$}
        \put(4,0){\matrixput(-18,-8)(8,0)3(0,8)4{\circle*2}
        \put(-2,0){\drawline(0,0)(-8,0)\drawline(-8,-8)(-8,8)
                \drawline(-16,16)(-8,0)}}
        \put(-8.7,-12.3){$z_0$}\put(4.5,-1){$z_1$}\put(-15,18){$z_2$}
        \put(-7,10){$w$}

        \put(10,9){$(\blue\l^2\red\m\black z_0\!:\!\blue\l\black z_1\!:\!
                        \blue\l\black z_2\!:\!\red\m\black w)$
                \normalsize$\in(\mathbb C^4\!-\!Z)/(\mathbb C^*)^2$}
        \put(20.5,0){$Z=\{z_0\!=\!w\!=\!0\}\cup\{z_1\!=\!z_2\!=\!0\}$}	\black
\EP
\\[20pt]			
\stepcounter{figure}						\makeatletter
\immediate\write\@auxout{\string\newlabel{Hirze}{\thefigure}} 	\makeatother
Fig. \thefigure: Hirzebruch surface $\mathbb F_2$ 
	and the blowup  
 	of $W\IP_{211}$.
\\[1pt]
\enddel 

\noindent{\bf Example 1: The Hirzebruch surface}
\\[2pt]
Hirzebruch surfaces $\mathbb F_n$ are $\IP^1$ bundels over $\IP^1$ that
can be defined by 
$	v_0=(0,-1),$ $ v_1=(1,0),$ $  v_2=(-1,n)$ and $ v_3=(0,1), 
$
with linear relations 
$v_0+v_3=0$, $nv_0+v_1+v_2=0$ and scaling parameters
$\m$ and $\l$ as shown in figure \ref{Hirze}.	

We consider the case $n=2$.
If we drop the vertex $v_3$ the fan would
consist of 3 cones and we obtain 
the weighted projective space
$W\IP_{211}$ with scaling weights $(2,1,1)$. This space is singular because 
the cone spanned by $(v_1,v_2)$ has volume 2. Indeed, if we drop the 
coordinate $w$ and set $\m=1$ then $(1\!:\!0\!:\!0)$ is a fixed \hbox{point} 
of the $\IC^*$ 
identification for $\l=-1$, i.e. we have a $\IZ_2$ quotient singularity.
The Hirzebruch surface $\mathbb F_2$ is a de\-singularization of this surface
corresponding to a subdivision of the cone $(v_1,v_2)$ into two basic
cones $(v_1,v_3)$ and $(v_3,v_2)$. The exceptional set is accordingly modified
to $Z=\{z_0=w=0\}\cup\{z_1=z_2=0\}$. We can now consider two cases:
If $w\neq0$ then $\m=1/w$ scales $w$ to $w=1$. This yields all
points $(z_0\!:\!z_1\!:\!z_2\!:\!1)\equiv(z_0\!:\!z_1\!:\!z_2)$ 
of $W\IP_{211}$ except for its singular point  $(1\!:\!0\!:\!0)$,
which is excluded due to the subdivision of the
cone $(v_1,v_2)$ by $v_3$. If $w=0$ we can scale $z_0\neq0$ to $z_0=1$
and find $(1\!:\!z_1\!:\!z_2\!:\!0)\in\mathbb F_2$. We thus observe that the
singular \hbox{point} has been replace 
by a $\IP^1$ with homogeneous coordinates 
$(z_1\!:\!z_2)$. This process of replacing a point by a projective space is 
called {\it blow-up}. In the present example it de\-singular\-izes a 
weighted projective space.
It can be shown \cite{Ful} that all singularities of toric varieties can 
resolved by a sequence of blow-ups that correspond to subdivisions of the 
fan.

\del
\def\ConeSigma{\drawline(0,10)(-20,40)\drawline(0,10)(20,40)
        \dashline2(0,10)(-5,50)\drawline(0,10)(5,30)
        \drawline(5,30)(20,40)\drawline(5,30)(-20,40)
        \drawline(-5,50)(20,40)\drawline(-5,50)(-20,40)}
\medskip\unitlength=1.2pt
\BP(100,55)
\put(80,0){\ConeSigma \putlab(-5,52)b{$z_1$}\putlab(22,42)b{$z_3$}
	\putlab(-22,42)b{$z_2$}\putlab(4,32)b{$z_0$}
	}	    
\put(20,0){\ConeSigma	\putlab(3,32)b{$\b$}\putlab(1,42)b{$\a$}
        \put(0,0){\blue\linethickness{2pt}\drawline(-20,40)(20,40)}\black}
\put(140,0){\ConeSigma	\putlab(9,47)b{$x$}\putlab(-14,46)b{$v$}
			\putlab(9,35)b{$u$}\putlab(-6,28)b{$y$}
        \put(0,0){\blue\linethickness{2pt}\drawline(-5,50)(5,30)}\black}
\EP
\\[-7pt]
\stepcounter{figure}						\makeatletter
\immediate\write\@auxout{\string\newlabel{ConiF}{\thefigure}} 	\makeatother
Fig. \thefigure: Toric desingularizations of the conifold.
\medskip
\enddel

\vspace{3pt}
\noindent{\bf Example 2: The conifold singularity}
\\[2pt]
According to theorem 2 the second source of singularities is 
nonsimplicity of a cone, which is only possible in at least 3 dimensions.
We hence consider a quadratic cone $\s$ as displayed in figure 
\ref{ConiF}b with generators
\\
$v_0=(1,0,0)$, $v_1=(0,1,0)$, $v_2=(1,0,1)$, $v_3=(0,1,-1)$
and relation $\sum q_iv_i=0$ with $q=(1,1,-1,-1)$ so that
\BE	\textstyle
	(z_0:z_1:z_2:z_3)=(\l z_0:\l z_1,\frac1\l z_2:\frac 1\l z_3),
	\label{ConiScale}
\EE
or $X=\IP(1,1,-1,-1)$.  
The dual cone $\s^\vee$ has generators
\\
\centerline{%
$m_0\!=\!(1,0,0)$, $\!m_1\!=\!(0,1,0)$, 
	$\!m_2\!=\!(0,1,1)$, $\!m_3\!=\!(1,0,-1)$.}
According to (\ref{Character})
the coordinate ring $A_\s$ 
is generated by
\vspace{-3pt}
\BEA
\HS0	x=\c^{m_0}=z_0z_2,\HS10 &&	y=\c^{m_1}=z_1z_3, \\\vspace{-4pt}
\HS0	u=\c^{m_2}=z_1z_2,\HS10 && 	v=\c^{m_3}=z_0z_3, \vspace{-2pt}
\EEA
which are regular, invariant under the scaling (\ref{ConiScale}) and
obey the relation $xy=uv$. Hence $A_\s=\IC[\s^\vee\cap M]	
\cong \IC[x,y,u,v]/\bra xy\!-\!uv\ket$ and $X\!=\!U_\s$ 
can be \hbox{identified} \hbox{with} the hypersurface $xy=uv$ in $\IC^4$,
which has a ``conifold singularity'' at the origin. 
%
As shown in
figure \ref{ConiF} the cone $\s$ can be triangulated in two different ways.
In the \hbox{first} case $\s_\a=\bra v_0,v_2,v_3\ket$, 
$\s_\b=\bra v_1,v_2,v_3\ket$ and the dual cones are
$	\s_\a^\vee=(m_\a,m_0,m_3),$ $\s_\b^\vee=(m_\b,m_1,m_2)
$
with $m_\a=-m_\b=(-1,1,1)$ so that we obtain the algebras
\BEA	\!A_\a=\IC(m_\a,m_0,m_3)\ni({z_1}/{z_0},\,x=z_0z_2,\,v=z_0z_3),&&
\\	\!A_\b=\IC(m_\b,m_2,m_1)\ni({z_0}/{z_1},\,u=z_1z_2,\,y=z_1z_3).&&
\EEA
With the exceptional set $Z=\{z_0=z_1=0\}$ we observe that the singular
point $x=y=u=v=0$ has been replaced by a $\IP^1\ni(z_0:z_1)$ and 
the transition functions show that $X_{\{\s_\a,\s_\b\}}$ can be identified 
with the total space of the rank two bundle $O(-1)\oplus O(-1)\to\IP^1$.
The reader may verify that the homogeneous coordinates work 
straight\-forwardly (geometrical)
in the sim\-pli\-cial cases (2a) and (2c). In the non-simplicial case (2b) 
the quotient (\ref{QuotRep}) has to be taken in the ``GIT--sense'' 
\cite{CoxHCR} because all $\IC^*$ orbits 
$(\l z_0:\l z_1:0:0)$ and $(0:0:\frac1\l z_2:\frac1\l z_3)$ map to the 
tip $0\in U_\s$ of the conifold (the {\it categorial quotient} of 
{\it geometric invariant theory} (GIT) involves the dropping of ``bad'' 
orbits).

The two 
toric resolutions 
correspond to blow\-ups replacing the singular 
point by two different $\IP^1$'s (which topologically are 2-spheres $S^2$).
We can hence go from one ``small resolution'' to the other via the
conifold by blowing down the $\IP^1$ to a point and blowing up that point
in a different way. This is called a {\it flop transition}. 

There is also a non-toric possibility to 
resolve the singularity 
by deforming the hypersurface equation to $xy-uv=\e$. 
Topologically this amounts to replacing the singularity by a 3-sphere $S^3$.
This can be seen as follows: By a linear change of variables 
$\{{x\mp y\02},{u\pm v\02}\}\leftrightarrow\{i^lw_l\}$
we can write the deformed conifold equation as 
$\sum_{l=1}^4 w_l^2=\e$. With $w_l=a_l+ib_l$ its real
and imaginary part 					become\vspace{-2pt}
\BE	
	\sum_{l\le4}a_l^2=\e+\sum_{l\le4} b_l^2,\HS30  \sum_{l\le4}
	a_lb_l=0.			
\EE								\vspace{-4pt}
For $\e>0$ the four real variables $a_k'=a_k/\sqrt{\e+\sum_l b_l^2}$ parametrize
a 3-sphere and the $b_k$ with $\sum_l a_l'b_l=0$ parametrize the fibers
of the cotangent bundle $T^*S^3$. The topology change between the small 
resolution and the deformation is called conifold transition. 

\del						****	work out the blow-up
Resolved conifold: $O(-1)\oplus O(-1)\to\IP^1$
\\[15pt]\phantom.\hfill
$\matrix{n_0&n_1&n_2&n_3\cr1&0&1&~0\cr0&1&0&~1\cr0&0&1&-1}$ ~ ~ ~
	$\matrix{x&y&z&w\cr1&0&0&~1\cr0&1&1&~0\cr0&0&1&-1}$ 
\\[30pt]$x=z_0z_2=t_1$, ~$y=z_1z_3=t_2$, ~$z=z_1z_2=t_2t_3$, 
\\$w=z_0z_3=t_1/t_3~~~\then~~~\IC[x,y,z,w]/\bra xy-zw\ket$ 
\\... unrelated to $q=(1,1,-1,-1)$.
\\[3pt]
$O(n)\to \IP^d$ ... homogeneous polynomials of degree $n$, transition 
functions: $P/(z_0^n)\to P/(z_1^n)$, 
$T\IP^1=O(2)$, three global sections
\\[3pt]Resolution of $\IP(1,1,-1,-1)$: triangulate $\to$ $\{[023],[123]\}$, 
$Z=\{z_0=z_1=0\}$,\BP(0,0)(-20,0)\putlin(0,-5,1,0,30)\putlin(30,-5,1,1,20)
\putlin(0,-5,1,1,20)\putlin(20,15,1,0,30)          \putlin(30,-5,-1,2,10)
\putlab(-3,-9)c{\scriptsize0}\putlab(33,-9)c{\scriptsize3}
\putlab(17,21)c{\scriptsize2}\putlab(50,21)c{\scriptsize1}
		\EP 
\\[5pt]two patches: $\cases{p^{02}_0=x=t_1,~q^{03}_0=w=t_1/t_3,
	~\l_0=t_1/t_2t_3={z_0\0z_1}&\cr
		p^{12}_1=z=t_2t_3,~~q^{13}_1=y=t_2,~~
	~\l_1=t_2t_3/t_1={z_1\0z_0}&\cr}$
\BP(0,0)\put(0,-9){$\then$ patch $x\to z\atop w\to y$ ... 
$\ex{z_1\0z_0}$}\EP
\\[9pt]transition: ``blow down'' / ``blow up'' ... 
\\{\bf holomorphic quotient:} $({z_0\0z_1},1,{z_2\0z_1},{z_3\0z_1})\sim
	(1,{z_1\0z_0},{z_2\0z_0},{z_3\0z_0})$ ... remove $Z\ex\IP^1~\iff$ 
	blow up $\IP^1$ \rlap{\tiny ... insert two $\IC^2$'s.}

\nibf Theorem: $X$ is projective if $\S$ is the normal fan of a polytope
	$\D\subset M_\IR$.
\\I.e. $\exists$ ample (halb/sehr ger\"aumig) line bundle $\to$ embedding 
in $\IP^N$; $\D\ni m~\to$ sections; {\bf monomial transition functions.}

A general toric variety has only mild singularities (being Cohen-Macaulay).
$m\in M$ gives a character $\c^m:T_N\to \IC^*$, regarded as a rational 
functions.						
\BE	\mao{div}(\c^m)=\sum_\r\bra m,v_\r\ket D_\r	
\EE
Fan $\S$ complete ($X$ compact) iff $|\S|=\IN_\IR$, orbifold ($V$-manifold)
iff $\S$ simplicial, non-singular iff $\IZ$-basis.

\bigskip\hrule
{\small	[ $\vec q$ integral or fractional if $v_j$ don't span $N$ 
		$\to$ finite quotient ...]}
\\The toric variety is characterized by {\bf allowed/identified limits} of
1-parameter subgroups!
$T$--{\bf orbits $\to$ fan}, which  $z_j$ may vanish simultaneously? 
$X\sim (\IC^r\minus Z)/(\IC^*)^{r-n}/G_{fin.ab.}$ with
$Z=\bigcup$(coord.planes) such that $z_{j_1}=\ldots=z_{j_s}=0$ only if
$n_{j_l}\in\s$ $\then$ $X=\bigcup U_\s$, other $z_j\to1$.
\\Example: $\IP^n$, ~ $\IP^a\ex\IP^b$: ~ ~ $t_i=z_i/z_0$, ~ $n_i=e_i$, 
	~ $n_0=-\sum e_i$, ~ ~ $\sum n_j=0~ \then~  q_j=1$.

\noindent
Regular functions on {\bf affine patch} $U_\s$: ~ $t^m$ with 
$\bra m,n_j\ket\ge0$ ... 
$U_\s=Spec(\IC[M\cap \s^\vee$]).
\\Spec=\{{\bf maximal} ideal=function zero on pt.\} $\subset$ 
\{{\bf prime} id.=irreducible set/Zariski 
\comment{~\\[-11pt]closed sets=algebraic sets}topology\}
\\[9pt]Example: $\IZ_n$ singularity:~
$n_1=\pmatrix{n\cr-1}, n_2=\pmatrix{0\cr1}$ ~$\then$~ $m_1=\pmatrix{1\cr0}$,
~ $m_2=\pmatrix{1\cr n}$
\\[9pt]$\IZ_2:~Spec\left(\IC[x=t_1,y=t_1t_2,z=t_1t_2^2]~/\bra xz=y^2\ket
	\right)=V(xz-y^2)	\subset\IC^3$, ~~~$y=\pm\sqrt{(xz)}$, 
\\$\IZ_n:\bra t_1,\ldots,t_1t_2^{n}
	\ket=\bra t_1^it_2^{n-i}\ket$
\\[9pt]
\nibf Theorem: $U_\s$ is nonsingular iff spanned by a lattice basis: 
	simplicial and volume=1
\enddel

\def\TwoDimRP{{
	\newcount\XYfac	\newcount\Yfac	\newcount\YA	\XYfac=10 \Yfac=18
	\newcount\XZfac	\newcount\Zfac 	\newcount\ZA	\XZfac=8  \Zfac=18
	\newcount\Xaux	\newcount\Yaux	\newcount\Zaux	\thicklines
		  \linethickness{2pt} 
\def\TwoDimPtGrid##1##2{{\matrixput
	      (-\CFigOffset,0)(\Yfac,0){##1}(0,\Zfac){##2}{\circle*5}}}
\def\Aa{\BP(60,60)(-10,-5)\blue
	\xyzline(0,0,0)(0,0,3)\xyzline(0,0,0)(0,3,0)\xyzline(0,0,3)(0,3,0)
	\black	\TwoDimPtGrid44 \EP}
\def\Ab{\BP(60,60)(-10,-5)\black
	\xyzline(0,0,0)(0,1,2)\xyzline(0,0,0)(0,2,1)\xyzline(0,1,2)(0,2,1)
	\black	\TwoDimPtGrid44	\EP}
\def\Ba{\BP(100,60)(-10,-5)\blue	
	\xyzline(0,0,0)(0,4,0)\xyzline(0,0,0)(0,2,2)\xyzline(0,2,2)(0,4,0)
	\black	\TwoDimPtGrid53	\EP}
\def\Bb{\BP(60,60)(-10,-5) \black
 	\xyzline(0,0,0)(0,2,0)\xyzline(0,0,0)(0,1,2)\xyzline(0,2,0)(0,1,2)
	\black	\TwoDimPtGrid33	\EP}
\def\Ca{\BP(60,60)(-10,-5) \blue		    \xyzline(0,0,0)(0,0,2)
	\xyzline(0,0,0)(0,2,0)\xyzline(0,2,0)(0,2,2)\xyzline(0,0,2)(0,2,2)
	\black	\TwoDimPtGrid33	\EP}
\def\Cb{\BP(60,60)(-10,-5) \black		    \xyzline(0,0,1)(0,1,0)
	\xyzline(0,2,1)(0,1,2)\xyzline(0,0,1)(0,1,2)\xyzline(0,1,0)(0,2,1)
	\black	\TwoDimPtGrid33	\EP}
\def\Da{\BP(60,60)(-10,-5) \black \xyzline(0,0,0)(0,0,2)\xyzline(0,0,0)(0,2,0)
	\xyzline(0,2,0)(0,2,1)\xyzline(0,0,2)(0,1,2)\xyzline(0,1,2)(0,2,1)
	\black	\TwoDimPtGrid33	\EP}
\def\Db{\BP(60,60)(-10,-5) \black \xyzline(0,0,1)(0,0,0)\xyzline(0,0,0)(0,1,0)
	\xyzline(0,2,1)(0,1,2)\xyzline(0,0,1)(0,1,2)\xyzline(0,1,0)(0,2,1)
	\black	\TwoDimPtGrid33	\EP}
\def\Ea{\BP(80,60)(-10,-5) 			    \xyzline(0,0,0)(0,0,2)
	\xyzline(0,0,0)(0,3,0)\xyzline(0,0,2)(0,1,2)\xyzline(0,1,2)(0,3,0)
	\black	\TwoDimPtGrid43	\EP}
\def\Eb{\BP(60,60)(-10,-5)
 	\xyzline(0,0,0)(0,1,0)\xyzline(0,2,1)(0,1,2)
	\xyzline(0,0,0)(0,1,2)\xyzline(0,1,0)(0,2,1)
	\black	\TwoDimPtGrid33	\EP}
\def\Fa{\BP(80,60)(-10,-5)			    \xyzline(0,0,0)(0,0,1)
	\xyzline(0,0,0)(0,3,0)\xyzline(0,3,0)(0,1,2)\xyzline(0,0,1)(0,1,2)
	\black	\TwoDimPtGrid43	\EP}
\def\Fb{\BP(60,60)(-10,-5)
 	\xyzline(0,1,2)(0,0,0)\xyzline(0,0,0)(0,2,0)\xyzline(0,2,0)(0,2,1)
	\xyzline(0,2,1)(0,1,2)
	\black	\TwoDimPtGrid33	\EP}

\def\SA{\BP(80,60)(-10,-5)
\red 	\xyzline(0,0,0)(0,1,2)\xyzline(0,0,0)(0,3,0)\xyzline(0,1,2)(0,3,0)
	\black	\TwoDimPtGrid43	\EP}

\def\SB{\BP(60,60)(-10,-5) 
\red 	\xyzline(0,0,0)(0,0,2)\xyzline(0,0,0)(0,2,0)
	\xyzline(0,2,0)(0,1,2)\xyzline(0,0,2)(0,1,2)
	\black	\TwoDimPtGrid33	\EP}

\def\SC{\BP(60,60)(-10,-5) 
\red 	\xyzline(0,0,0)(0,0,1)\xyzline(0,0,0)(0,2,0)
	\xyzline(0,2,0)(0,2,1)\xyzline(0,0,1)(0,1,2)\xyzline(0,1,2)(0,2,1)
	\black	\TwoDimPtGrid33	\EP}

\def\SD{\BP(60,60)(-10,-5) 
\red 	\xyzline(0,1,0)(0,0,1)\xyzline(0,1,0)(0,2,0)\xyzline(0,0,2)(0,0,1)
	\xyzline(0,0,2)(0,1,2)\xyzline(0,2,0)(0,2,1)\xyzline(0,1,2)(0,2,1)
	\black	\TwoDimPtGrid33	\EP}
\parbox{16cm}
{\begin{center}\unitlength=.8pt      {\Yfac=12 \Zfac=12 \hspace{-3pt}
        \Aa ~ \Ab} ~~~~ \Ba ~ \Bb ~~~~~ \Ca ~ \Cb      \\[3pt]
        \Da ~ \Db ~~~~~ \Ea ~ \Eb ~~~~~ \Fa ~ \Fb       \\[3pt]
        \SA ~~~~~~~ \SB ~~~~~~~ \SC ~~~~~~~ \SD         \end{center}}
}}


\vspace{-5pt}
\subsection{Line bundles}
\vspace{-2pt}
The conifold is 
the standard example of a {\it non-compact} (local) Calabi--Yau geometry. 
A {\it compact} toric varieties, on the other
hand, never have $c_1=0$.
We will hence not only be interested in toric varieties themselves but
also in hypersurfaces or complete intersections thereof, which under 
appropriate conditions are smooth Calabi--Yau spaces. Their defining equations
\hbox{will} be sections of non-trivial line bundles.
The relevant data of these bundles is the transition functions between 
different patches. This data is closely related
to the topological data of 
{\it Cartier divisors}, which by definition are locally given in terms of
rational equations $f_\a=0$ with $f_\a/f_\b$ regular and nonzero on
the operlap of two patches. Sin\-ce multiplication by a rational function does
not change the line bundle we are interested in {\it divisors classes}
with respect to {\it linear equivalence}, i.e. modulo addition of 
{\it prin\-ci\-pal divisors} $\mao{div}(f)$, which are the divisors of 
rational {\it functions} $f$. 
Cartier divisor {\it classes} hence 
determine the Picard group $\mao{Pic}(X)$ of holomorphic line bundles. 

Finite formal sums of irreducible varieties of co\-dimen\-sion one are 
called {\it Weil divisors} (which may not be Cartier, i.e. locally principal,
on singular varieties). On a toric variety it can be shown that 
the {\it Chow group} $A_{n-1}(X)$ of Weil divisors modulo linear 
equivalence is generated by the $T$-invariant irreducible divisors 
$D_j$ modulo the 
principal divisors $\mao{div}(\c^m)$ with $m\in M$, i.e. there is an
exact sequence
\BE	0\rightarrow M\rightarrow \IZ^{\S^{(1)}}\rightarrow A_{n-1}(X)
	\rightarrow0
\EE
where $M\ni m\to (\bra m,v_j\ket)\in\IZ^{\S^{(1)}}$ and 
$\IZ^{\S^{(1)}}\ni (a_j)\to
\sum a_j D_j$, The Chow group $A_{n-1}(X)$ hence has rank $r-n$.
It contains the Picard group as subgroup, which is torsion free if $X_\S$ is 
compact \cite{Ful}.

A Weil divisor of the form $D=\sum a_j D_j$ is Cartier, and hence defines a 
line bundle ${\cal O}(D)\in \mao{Pic}(X)$, if there exists an $m_\s\in M$ for 
each maximal cone $\s\in\S$ such that $\bra m_\s,v_j\ket=-a_j$ for
all $\r_j\in\s$. The transition functions of ${\cal O}(D)$ between the 
patches $U_\s$ and $U_\t$ are then given by $\c^{m_\s-m_\t}$.
If $X$ is smooth then all Weil divisors are Cartier. For a
simplicial fan 
$kD$ is Cartier for some positive integer $k$.
%
For Cartier divisors 
the $\S$-piecewise linear real function $\ps_D$ on $N_\IR$ defined by
\BE	\ps_D(v)=\bra m_\s,v\ket \HS20 \mathrm{for} \HS20 v\in\s
\EE 
is called support function. 
\del					****
\\$X$ complete $\then~\mao{Pic}(X)$ is torsion free, simplicial $\then$ 
finite index in the Chow group, smooth then $\mao{Pic}(X)=A_{n-1}(X)$.
\\
\enddel
If $X$ is compact and $D=\sum a_j D_j$ Cartier 
then ${\cal O}(D)$ is generated by global sections iff the 
support function $\ps_D$ is  
convex and $D$ is ample iff $\ps_D$ is strictly convex, 
i.e. if $\bra m_\s,v_j\ket>-a_j$
for $\dim\s=n$ and $\r_j\not\subset\s$.
For convex support functions
\BEA	\label{DelD}
	\D_D&=&\{m\in M_\IR:\bra m,v_j\ket\ge-a_j ~~~~\forall ~~j\le r\}
\\	&=&\{m\in M_\IR:\bra m,u\ket\ge\ps_D(u)~~ \forall ~u\in N\}
\EEA
defines a convex lattice polytope $\D_D\subset M_\IR$ whose lattice points
provide the global sections of the line bundle ${\cal O}(D)$ 
corresponding to the divisor $D$
(the first equality defines $\D_D$ also if $D$ is not Cartier). In particular,
$\D_{kD}=k\D_D$ and $\D_{D+\mao{div}(\c^m)}=\D_D-m$ so that the polytope
can be translated in the $M$ lattice without changing the divisor class and
the transition functions.			

In terms of the polytope (\ref{DelD}) $D$ is generated by
global sections iff $\D_D$ is the convex hull of $\{m_\s\}$ and $D$ is
ample iff $\D_D$ is $n$-dimensional with vertices $m_\s$ for $\s\in\S^{(n)}$ 
and with $m_\s\neq m_\t$ for $\s\neq\t\in\S^{(n)}$.
In the latter case there is a bijection between faces of $\D_D$ and 
cones in $\S$ or, more pricisely, $\S$ is the {\it normal fan} of $\D_D$:
By definition the cones 
$\s_\t$ of the normal fan $\S_\D$ of a polytope $\D$ are the dual cones 
of the cones over $\D-x$ where
$x\in M_\IR$ is any point in the relative interior of a face $\t\subset\D$.
If $0\in M$ is in the interior of $\D$, as can always be achieved by a 
rational translation of $\D$ by $\d x\in M_\IQ$, then the normal fan $\S_\D$ 
coincides with the fan of cones over the faces of the {\it polar}
polytope $\D^\circ\subseteq N_\IR$ defined by
\BE							\label{polar}
	\D^\circ=\{y\in N_\IR~:~~\bra x,y\ket\ge -1~~\forall\, x\in \D\}.
\EE
On smooth compact toric varieties it can be shown that every ample 
$T$-invariant divisor is very ample. The sections $\c^m$ of 
such an 
${\cal O}(D)$ hence provide an embedding 
of $X_\S$ into $\IP^{K-1}$ via $(\c^{m_1}:\ldots:\c^{m_K})$ 
where $K=|\D_D\cap M|$ 
is the dimension of the space of global sections of  ${\cal O}(D)$. 
A toric variety $X_\S$ therefore is {\it projective} iff $\S$
is the normal fan of a lattice polytope $\D\subset M_\IR$ \cite{Ful,Oda}.

Summarizing, the equations defining Calabi--Yau
hypersurfaces or complete intersections will be 
sections of 
line bundles ${\cal O}(D)$ 
given by Laurent polynomials 
\BE	f=\sum_{m\in\D_D\cap M}c_m\c^m=\sum_{m\in\D_D\cap M}c_m\,
		\prod_j z_j^{\bra m,v_j\ket}
\EE
whose exponent vectors $m$ span the convex lattice polytopes 
$\D_D\subseteq M_\IR$ defined in eq. (\ref{DelD}). In an affine patch 
$U_\s$ the local section $f_\s=f/\c^{m_\s}$ is a regular function 
$f_\s\in A_\s$ because $\D_D-m_\s\subset\s^\vee$.

\vspace{-5pt}
\subsection{Intersection ring and Chern classes}
\vspace{-2pt}
If a collection $\r_{j_1},\ldots,\r_{j_k}$ of rays is not contained in a 
single cone then the corresponding homogeneous coordinates $z_{j_l}$ are not 
allowed to vanish simultaneously and the corresponding divisors $D_{j_l}$
have no common intersection. For the intersection ring we hence expect
the {\it non-linear relations} $R_I=D_{j_1}\cdot 	
	\ldots\cdot D_{j_k}=0$, 
where it is sufficient to take into account the {\it primitive collections} 
$I=\{j_1\ldots j_k\}$ as defined by Batyrev, i.e. the {\it minimal} index 
sets such that the corresponding rays do not all belong to the same cone 
(cf. the definition of the exceptional set $Z=\bigcup Z_I$ in section
\ref{HomoCoord}). 
The ideal in $\IZ[D_1,\ldots, D_r]$
generated by these $R_I$ is called Stanley--Reisner ideal $J$, and 
$\IZ[D_1,\ldots, D_r]/J$ is the Stanley--Reisner ring.

The Chow groups $A_k(X)$ of a variety $X$ are generated by $k$-dimensional 
irreducible closed subvarieties of $X$ modulo rational equivalence by 
divisors of rational functions on subvarieties of dimension $k+1$.
For an arbitrary toric variety $X_\S$ it can be shown that $A_k(X)$ is 
generated by the equivalence
classes of orbit closures $V_\s$ for cones $\s\in\S^{(n-k)}$.
The intersection ring of a {\it non-singular compact} toric variety $X_\S$
is \cite{Dan}  
\BE								\vspace{-3pt}
	A_*(X_\S)=\IZ[D_1,\ldots, D_r]\,/\bigl\bra R_I,
		\hbox{$\sum_j$}\bra m,v_j\ket D_j\bigr\ket	\vspace{-3pt}
\EE
(for a definition of the intersection product see \cite{Ful}). 
The intersection ring 
can hence be obtained from the Stanley--Reisner ring by adding 
the {\it linear
relations} $\sum_j\bra m,v_j\ket D_j\simeq 0$, where it is sufficient to take
for $m$ a set of basis vectors of the $M$-lattice. 
The Chow \hbox{ring} also determines the homology groups 
$H_{2k}(X_\S,\IZ)=A_k(X,\IZ)$. These results actually generalize to the 
{\it simplicial projective} case with the 
exception that one needs to admit rational coefficients~\cite{Oda,Ful}.
In particular, for a maximal-dimensional simp\-licial cone $\s$ spanned by 
$v_{j_1},\ldots, v_{j_n}$ the intersection number of the corresponding
divisors is
\BE							\vspace{-3pt}
	D_{j_1}\cdot\ldots\cdot D_{j_n}=1/\mao{Vol}(\s)	\vspace{-3pt}
\EE
where $\mao{Vol}(\s)$ is the lattice-volume (i.e. the geometrical volume 
divided by the volume $1/n!$ of a basic simplex). 

Having discussed the cycles we not turn to differential forms.
The canonical bundle of a non-singular toric variety can be obtained by 
considering the rational form
\BE								\vspace{-3pt}
	\o=\frac{dx_1}{x_1}\wedge\ldots\wedge\frac{dx_n}{x_n}	\vspace{-3pt}
\EE
which by an appropriate choice of the orientation (i.e. the order 
of the local coordinates $x_i$ in the affine patches) is a rational 
section of $\O_X^n$. This implies 
\BE
	\O_X^n={\cal O}_X(-\textstyle\sum\limits_{j=1}^r D_j)
\EE
and for the canonical divisor $-D=-\sum D_j$. The computation of the total
Chern class requires an expression for the (co)tangent bundle, for which
there is an exact sequence
\BE	\textstyle
	0~\to~\O_X^1~\to~\O_X^1(\log D)~\rel {\mao{res}} 
	\to~\bigoplus_j{\cal O}(D_j)
\EE
where $\O_X^1(\log D)$ turns out to be trivial and the residue map takes 
$\o=\sum f_j\,dz_j/z_j\rel {\mao{res}} \to\oplus f_j{}_{|_{D_j}}$ \cite{Ful}. 
A calculation yields the total Chern class of the tangent bundle 
\BE	\textstyle
	c(T_X)=\prod_1^r(1+D_j)=\sum_{\s\in\D}[V_\s]
\EE
and the Todd class
\BE	\textstyle
	\mao{td}(T_X)=\prod_1^r\frac{D_j}{1-\exp(-D_j)}=1+\frac12 \,c_1
			+\frac1{12}\,c_1^2c_2+\ldots
\EE
The first Chern class $c_1=\sum D_j$ is positive for compact
toric varieties, but it vanishes for the 
conifold because of the
linear relations $	
			D_0+D_2\sim0$, 
	$D_1+D_3\sim0$ and $D_1\sim D_3$.
(Implications for volumes and numbers of lattice points 
can now be derived by applying the 
Hirzebruch--Riemann--Roch formula 
$
	\c(X,E)=\int\mao{ch}(E)\mao{Td}(X)
$
to the case of line bundles of Cartier divisors
as described, for example, in the last chapter of \cite{Ful}.)

\vspace{-5pt}
\subsection{Symplectic reduction}
\vspace{-2pt}
There is another approach to toric geometry in terms of symplectic 
instead of complex geometry, which is important because, in addition 
to the complex structure, we will also need a K\"ahler metric. Moreover,
the symplectic approach can be given a direct physical interpretation in terms
of supersymmetric gauged linear sigma models \cite{phases}.
The idea is that the $\IC^*$--quotient can be performed in two steps:
We first divide out the phase parts, which amount to compact $U(1)$
quotients, and then -- instead of a radial identification -- fix the 
values of appropriate ``radial'' variables
to certain sizes $t_a$ that will parameterize the K\"ahler metrics.

In order that the quotient inherits a K\"ahler form (and hence a 
symplectic structure) from the natural K\"ahler form on  $\IC^r$
\BE	\textstyle
	\o=i\sum dz_j\wedge d\bar z_j=2\sum dx_j\wedge dy_j=\sum
	dr_j^2\wedge d\ph_j, 					\label{Kform}
\EE
with $z=x+iy=re^{i\ph}$ 
we use the symplectic re\-duct\-ion formalism. This requires that the 
$G$--action is hamil\-tonian,
i.e. given by a moment map $\m:\IC^r\to \eg^*$ to the dual $\eg^*$
of the Lie algebra $\eg$ of $G=U(1)^{r-n}$ such that the 
hamiltonian flows defined by
$\m$ generate the infinitesimal $G$-transformations. Then the symplectic
reduction theorem guarantees that the restriction of the image of the 
moment map to fixed values $t_a$ for $a=1,\ldots,n-r$  
induces a symplectic structure on the quotient of the preimage
$\m^{-1}(t_a)/G$ for regular values of $t_a$. 

In toric geometry we consider the moment maps 
\BE	\textstyle						\label{MM}
	\m_a=\sum_jq_j^{(a)}|z_j|^2 \qquad\hbox{with}\quad\sum_jq_j^{(a)}v_j=0
\EE
With
$\o^{-1}\sim\sum_j\frac\6{\6\ph_j}\wedge\frac\6{\6r^2_j}$ the corresponding
hamiltonian flows $\o^{-1}(\m_a)\sim\sum_jq_j^{(a)}\frac\6{\6\ph_j}$ 
generate the compact subgroups of the $\IC^*$ actions of the homomorphic 
quotients (\ref{QuotRep}).
In the gauged linear sigma model \cite{phases} the $q_j^{(a)}$ are the 
charges of $r$ chiral superfields $z_j$ under a $U(1)^{r-n}$ gauge group 
and the moment maps $\m_a$ are $D$--terms in the superpotential. The imaginary
parts of the complexified radii $t_a$ thus correspond to $\th$-angles.

Under the symplectic reduction the holomorphic $r$-form $\O=\prod dz^j$ 
on $\IC^r$ descends to a holomorphic $n$-form iff it is invariant under
the group action, i.e. if $\sum_j q_j^{(a)}=0$ for all $a\le r-n$. 
These equations can be interpreted as $U(1)$ gauge anomaly cancellation 
conditions in the linear sigma model \cite{phases}. Since the existence of a 
holomorphic $n$-from on $X_\S$ is equivalent to $c_1=0$ we thus obtain
a simple form of the Calabi--Yau condition (cf. the vanishing of 
$\sum_j q_j$ for the conifold).

Instead of the linear combinations (\ref{MM}) we can consider all 
moment maps $\m_j=|z_j|^2$, whose flows are phase rotations 
of the homogeneous coordinates $z_j$. After the symplectic reduction the 
effective part ${\cal G}\cong U(1)^n$ of this $U(1)^r$-action yields the 
compact part of the torus action ${\cal G}\subset T$. The image of the 
corresponding momentum map is a convex polytope, the {\it Delzant 
polytope}  $\D(t_a)$, whose corners correspond to fixed points of $\cal G$.
Since the Lie algebra $\eg$ of $\cal G$ can be identified with the 
the real extension of the $N$-lattice $\D(t_a)$ is a polytope in 
$\eg^*\equiv M_\IR$.
\\[3pt]
{\bf Example 3:} For the projective space $\IP^n$ all $q_j=1$ and
with $t_a=r^2$ we obtain the symplectic quotient as 
$\{|z_0|^2+\ldots+|z_n|^2=r^2\}/U(1)$. The value $t_a=r^2$ of the 
moment map of the symplectic reduction hence parametrizes the size of the 
projective space.
The image of the moment map for ${\cal G}\subset T$ on the resulting
toric variety is 
the simplex $\{t_i\ge0,r^2-\sum_1^n t_i=t_0\ge0\}\subset M_\IR$ whose faces 
of codimension $k$ correspond to the vanishing of $k$ moment maps $t_i$
and hence to fixed points of a $U(1)^k$ subgroup of $\cal G$. 
The fan of the toric variety $\IP^n$ is the normal fan of $\D$.
We thus can construct $\IP^n$ as a (compact) torus fibration over a 
polytope $\D\in\IR^n$ whose fibers degenerate to lower-dimensional 
tori over the faces of $\D$.
This fibration structure has been used by Strominger, Yau and Zaslow 
\cite{SYZ} for an interpretation of mirror symmetry as
$T$-duality on a torus-fibered Calabi--Yau manifold.
\\[3pt]
{\bf Example 4:} As a non-compact example we consider the conifold whose
K\"ahler metric is parametrized by $t=|x_0|^2+|x_1|^2-|x_2|^2-|x_3|^2$. 
Obviously $t=0$ is a singular value, while for $t=\pm \e^2\to0$ the size $\e$
of one of the blown-up $\IP^1$'s shrinks to 0. Regular values
of the moment maps $t_a$ lead to a smooth symplectic quotient 
and, in particular, to a (projective) triangulation of the fan. The 
corresonding smooth K\"ahler metric is parametrized by the $r-n$ values $t_a$,
which can be interpreted a sizes of certain two-cycles, in accord \hbox{with}
the dimension $r-n$ of $H_2(X_\S)$. The regular values correspond to open
cones of the {secondary fan}, \hbox{which} parametrizes the K\"ahler moduli 
spaces and whose chambers are separated by walls that correspond to {\it flop 
transitions} \cite{GKZ} between different smooth {\it phases} 
(in the physicist's language \cite{phases}). At the transition a cycle 
shrinks to a point that is blown up according to a different triangulation 
$\S_\D(t)$ on the other side of the wall \cite{GKZ}.

\del
Theorem (Delzant polytope): The image of the momentum map is convex.
The momentum maps for the torus action of a toric variety $X$ 
yields an projection $X\to \D_D\in M_\IR$ whose {\bf fibers} are $T^l$ with 
$l$={\bf dimension of torus orbit}=dimension of the face. 
$\S$ is the normal fan.

The moment map can be interpreted as a map from the toric
variety $X$ to the $M$-lattice with torus fibers so that we can visualize 
a toric variety as a torus fibration over a convex polytope with the
fibers degenerating over the faces of the polytope in a way that is 
given by the combinatorial data of the fan. Such a fibration has been used
by Strominger, Yau and Zaslow for an interpretation of mirror symmetry as
$T$-duality on a torus-fibered Calabi--Yau manifold. The symplectic approach
is also very useful for the construction of special lagrangian submanifolds,
which provide the support of $D$-branes on Calabi--Yau manifolds and hence
for the generalization of mirror symmetry to open strings \cite{AgaVafa}.
\enddel

\del
In \cite{AgaVafa} the GLSM is used to construct SLAG manifolds 
(lagrangian submanifolds where the restriction of $\O$ has a constant phase)
in the non-compact case (using a slightly modified moment map for which
the fibers are non-compact).

\noindent
Moreover: $\D$ is the basis of a Lagrangian fibration with $T^n$ fibers
(e.g. $\IP^1=S^1\to I$)
\\[5pt]
\nibf Definition: A {\bf Hamiltonian $G$-manifold} is a triple $(M,\o,\m)$ 
consisting of an oriented $G$-manifold $M$ (i.e., a manifold with an 
action $\t_g$ of a Lie Group $G$ with $\t_{gh}=\t_g\t_h$), a $G$-invariant 
symplectic form $\o$ and a {\bf moment{\rm um} map} $\m:M\to\eg^*$, i.e. a map
implementing infinitesimal group actions $\eg\ni\x\to \x^M\in TM$ as 
{\bf Hamiltonien flows} $X_\m$ with {\tiny $\o(X_\m,df)=\th(\m,f)$ }
\BE     d\m^\x=i_{_{\x^M}}\o ~~~~\mathrm{ with } ~~~~
        \m^\x(x)=\langle\m(x),\x\rangle
\EE
that is equivariant w.r.t. the $G$ action
on $M$ and the coadjoint action on $\eg^*$, i.e. 
$\m\circ g= \mao{ad}_g^*\circ \m$.
\\$\m$ ... $1^{st}$ class constraints: $[X_f,X_g]=X_{\{g,b\}}$ ...
$\m\equiv f,g$ constant along $X_\m$.
\\[5pt]
\nibf Example: The circle action $\x=\sum q_i \6_{\th^i}$ with 
weights $q_i$ on $\IC^n$ with the standard symplectic form 
$\o=\sum dx^i\wedge dy^i$ has moment map $\m=-\2\sum q_jr_j^2$ where
$\6_\th=x\6_y-y\6_x$ and $r^2=x^2+y^2$. 
\\[3pt]
$\bullet$ $T^2=\IR^2/2\p\IZ^2$ with $S^1$ action $x\to x+\th$ has no
moment map because $d\m=i_{\6_x}dx\wedge dy=dy$ has no global/periodic 
solution.  //Locally Hamiltonian flow $\not\then$ global; Toric: $T=\eg\to
	M_\IR=\eg^*$.

\newpage
\nibf Theorem (symplectic reduction / Marsden-Weinstein): 
If $t\in \eg^*$ is a regular value {\small with free + proper $\mao{Ad}^*$
action of the stabilizer on $\eg^*$}
of {\footnotesize an $\mao{Ad}^*$-equivariant}
momentum map $\m$ then $X=\m^{-1}(t)/G$ has a unique symplectic form 
$\o_\m$ with $\p^*\o_\m=i^*\o$, where 
$M \rel i \leftarrow \m^{-1}(t) \rel {\pi} {\rightarrow} X$.
\\{\small 
}
\nibf Toric: $-\m^A=\sum q_i^A|z_i|^2=t^A$, $A\le r-n$ K\"ahler 
parameters/sizes of divisors dual to $h_{11}$
\\(Chow group $A_{n-1}=G^*$)
\\[3pt] 
\nibf Theorem (Delzant polytope): The image of the momentum map is convex.
The momentum maps for the torus action of a toric variety $X$ 
yields an projection $X\to \D_D\in M_\IR$ whose {\bf fibers} are $T^l$ with 
$l$={\bf dimension of torus orbit}=dimension of the face. 
$\S$ is the normal fan.
\\Example: $\IP^1$, line segmet = $\{t_i\ge0,t_1+t_2=r^2\}$, $\IP^2$, ...
(integral iff the symplectic structure comes from an ample divisor).

\nibf Calabi-Yau: $\O=\prod dz^j$ descends to a holomorphic $n$-form iff it
is invariant, i.e. $\sum_j q_j^A=0~~\forall A$ \\ $\iff~~c_1=0~~\iff~$  
anomaly cancellation in GLSM.
\\[5pt]
Conifold transition (flop): $t=|x_0|^2+|x_1|^2-|x_2|^2-|x_3|^2$, singular
at $t=0$ ($\o_{il}\sim g_{ij}$ degenerate).

\nibf A/B-type branes: 4d-SUSY $\then$ N=2 SCFT 
[Banks, Dixon, Friedan, Martinec, 1988]
\\[2pt]
(X,Y)$_{real}\to (Z,\5Z)$ ... $G=G^++G^-$ ${G^+,G^-}=T+J$, $[J,G^\pm]=G^\pm$
\\[-7mm]\phantom.\hfill$\cases{~J_L=g_{\m\nb}\ps_L^\m\ps_L^\nb
		\rlap{$\sim\6\Ph_L$}&\cr
	G_L^+=g_{\m\nb}\ps_L^\m\6X^\nb&\cr
	G_L^-=g_{\m\nb}\6_X^\m\ps_L^\nb&\cr}\!\!\!\!\!\!\!\!\!\!\!\!\!$
\\[-6mm]{}[Ooguri,Oz,Yin, th/9606112]: keep SUSY$\sim$spectral flow 
$e^{i\Ph_L}=\O_{\m \ldots\r}\ps^\m\ldots\ps^\r$
\\A-type b.c.: $J_L=-J_R$, ~$G_L^+=\pm G_R^-$, ~ $e^{i\Ph_L}=e^{-i\Ph_R}$
\\B-type b.c.: $J_L=J_R$, ~~~$G_L^+=\pm G_R^+$, ~
	$e^{i\Ph_L}=(-)^ne^{i\th}e^{-i\Ph_R}$
\\both preserve $N=1$ SCFT with $T_L=T_R$, $G_L=\pm G_R$.

\noindent
b.c. for fields: $\6X^i=R^i{}_j\5\6X^j\atop \ps_L^i=\pm R^i{}_j\ps_R^j$ ...
$R$ orthogonal: $R^TgR=g$, ~symmetric $\to$ EV
$=\pm1\to\cases{\!\!\!\!&Neumann\cr\!\!\!\!&Dirichlet\cr}$
\\[3pt]non-symmetric $\sim$ gauge field strength $\sim$ mixed b.c.

\nibf B-type: $R^T\o R=\o~~\then$ mixed $\o_{tang,orth}=0\then$ dim$\in2\IZ$, 
~~$\O R\ldots R=e^{i\th}\O~~~\then e^{i\th}=(-)^{d-p/2}$

\nibf A-type: $R^T\o R=-\o~~\then$ $\o$ blockdiagonal=Lagrangian $d\02$-cycle 
~$\O R\ldots R=\5\O~\to\mao{Im}\O=0$ SLAG 
special LAG [Harvey, Lawson 85]::calibrated geometry / minimal volume
~~~\rlap{\footnotesize
	$\hbox{Kapustin th/0311101}\atop\hbox{ MS-Bibel p.800-813}$}

\nibf Branes in toric geometry {\rm th/0012041}: 
	{\bf Mirror symmetry = T-duality} of $S^3$ fiber,
\\brane along fiber: Neumann=wrapped $\leftrightarrow$ Dirichlet=fixed. 
\\[3pt]
K\"ahler form $\o=i\sum dz^id\5z^i=\sum d|z^2|\wedge d\th$, 
\\LAG, assume rational slope in $T^n$ ... $\sum q_i^\a|z^i|^2=c^\a$
\\kernel $q^\a v_\b=0$ $~\then~$ $\sum v_\b^i\th^i=0$, $\th^i=q_i^\a\Ph_\a$
$~\then~$ middle dimensional,
\\SLAG: $\O$ has constant phase, given by $\sum \th^i$, hence some
$v=(1\ldots1)~~\to$
$\sum_i q_i^\a=0~~\forall \a$.
\\LAG w.r.t. wrong metric, but flow to ... in the IR

\thinlines
\BP(300,80)(-200,-30)   \putlab(-200,10)l{$A$-branes: $S^1\ex\IR_+^2$}
                        \putlab(-200,-10)l{$B$-branes: $\IP^1$}
        \putlin(-30,0,1,0,60)\putlin(-30,0,-1,1,20)\putlin(-30,0,-1,-1,20)
        \putlin(30,0,1,1,20)\putlin(30,0,1,-1,20)\putlin(0,0,1,2,10)
        \putlab(20,20)b{$S^1\ex\IR_+^2$}
        \putlab(100,10)l{local sing. $xz=f(y,w)=yw+O(3)$}
        \putlab(100,-10)l{ ~ ~ ~ like $f(y,w)=(1-e^y)(1-e^z)$}
\EP
\bigskip\hrule\bigskip

$\IR\to T^2$ fibration: 
$\o=i dz\wedge d\5z=2dx\wedge dy=d r^2\wedge d\ph~\then~\th
	=i\6_z\wedge\6_{\5z}=\2dy\wedge dx=\6_\ph\wedge\6_{r^2}$\\
$t_\a=|z_1|^2-|z_3|^2,~~t_\b=|z_2|^2-|z_3|^2$,~
$t_\g=\mao{Im} z_1z_2z_2~\then~ \d_\g z_i=\2(\5z_1\5z_2\5z_2/\5z_i), 
	~\d_\g \5z_i=\2(z_1z_2z_2/z_i)$

McLean '92: $\dim(deform.(L))=b_1(L)$ $\to$ SYZ '96
\bigskip\hrule\bigskip

D-branes: Allgemein $\sum Q_i^A|z_i|^2=t^A$, ~~~~$\sum Q_i^A=0$
\\$|z_1^2|+|z_2^2|-|z_3^2|-|z_4^2|=t$
\\$r_\a=0,r_\b=r_1^*,r_\g\ge0$~~~~~~~~~~~ $U_1=(z_1,z_2,z_3)$,
\\$r_\b=0,r_\a=r_2^*,r_\g\ge0$~~~~~~~~~~~ $U_1=(z_1,z_2,z_4)$,
\\$r_\a-\r_\b=0,r_\a=r_3^*,r_\g\ge0$

$O(-3)\to\IP^3$ $|z_1^2|+|z_2^2|+|z_3^2|-3|z_0^2|=t$

$r_\a=|z_1^2|-|z_2^2|$ $r_\b=|z_2^2|-|z_4^2|$
$2|z_2^2|+|z_1^2|-|z_3^2|-t$

$\o_{|_L}=0$, $\mao{Im}(e^{i\th}\O_{|_L}=0$ $Vol(L)=\int_L\O$ minimal
\\(-q,p)-Zykel entartet №ber der Kante in $\G$, die
zu $p r_\a+q r_\b=0$ gehЎrt.

\enddel

\begin{figure*}[htb]					\begin{center}
\vspace*{-18pt}\TwoDimRP	\vspace*{-25pt}
\stepcounter{figure}						\makeatletter
\immediate\write\@auxout{\string\newlabel{TwoDimRP}{\thefigure}} \makeatother
\\[25pt]{Fig. \thefigure: All 16 reflexive polygons in 2D: 
	The first 3 dual pairs are maximal/minimal and
\\	contain all others as subpolygons, while the last 4 polygons are 
	selfdual.~~}					\end{center}
\vspace*{-19pt}\end{figure*}


\section{Strings, geometry and reflexive polytopes}	\label{Sec:CY}

At observable energy scales string theory leads to an effective
theory that corresponds to a 10-dimensional supergravity compactified
on a 6-dimensional manifold $K$. At small distances space-time hence
looks like $M_4\ex K$, where $M_4$ is our 4-dimensional Minkowski space,
as long as quantum fluctuations of the metric are sufficiently small to allow
for a semiclassical geometrical interpretation.

For phenomenological reasons we usually require that supersymmetry
survives the compactification, \hbox{which} implies the existence of a 
covariantly
constant spinor $\nabla \h=0$ on the internal manifold $K$.
In the simplest situation RR background fields and the $B$ field
vanish. Candelas, Horowitz, Strominger and E.~Witten \cite{CHSW} showed 
that this implies that $K$ is a complex K\"ahler manifold with vanishing
first Chern class. 
(The inclusion of $B$ fields was already discussed in a beautiful paper
by Strominger \cite{Strom}, but RR fluxes were largely for a long
time until their importance for moduli stabilization in type II theories
was recognized \cite{Kachru}. The investigation of their geometry 
lead to the beautiful new concept of generalized complex structures 
\cite{Hitchin,Gualt,Grana}.) Explicitly, the K\"ahler form $\o$ and the 
holomorphic 3-form $\O$ of the Calabi--Yau can be constructed in terms 
of 
$\h$ as
\BE
	\o_{ij}=i\h^\hc\g_{[i}\g_{j]}\h, \HS20
	\O\sim\h^\hc\g_{[i}\g_j\g_{k]}\h,
\EE
and the integrability condition $N_{ij}{}^k=0$ for the complex structure 
$J_i{}^l=\o_{ij}g^{jl}$
with the Nijenhuis tensor
$
	N_{ij}{}^k=J_i{}^l\6_l J_j{}^k-J_j{}^l\6_l J_i{}^k
                -\6_iJ_j{}^l J_l{}^k+\6_jJ_i{}^l J_l{}^k
$
is a trivially satisfied for the torsion--free metric-compatible connection 
that stabilizes $\h=0$.

The condition $c_1=0$, which is equivalent to the existence of a 
holomorphic 3-form $\O$, has been conjectured by Calabi 
and proven by Yau to be also equivalent to the existence of a Ricchi-flat 
K\"ahler metric, so that the vacuum Einstein equations are satisfied.

\del
, ``Vacuum
Proposal.tex:P

Nienhuis tensor $N_{ij}{}^k==J_i{}^l\6_l J_j{}^k-J_j{}^l\6_l J_i{}^k
                -\6_iJ_j{}^l J_l{}^k+\6_jJ_i{}^l J_l{}^k$

{\bf Hermitian metric} $g(JX,JY)=g(X,Y)$ $\to$ {\bf K\"ahler form}
$\o(X,Y)\equiv g(JX,Y)=-\o(Y,X)$ {\bf exists $\forall$ compl.MF:} 
$\2(g(X,Y)+g(JX,JY))$ ~~ 

Hermitian connection $Dg=DJ=0$, {\bf torsion free$\equiv$K\"ahler MF}: 

Calabi-Yau/SUSY$\!\!_{_{_{4d}}}\!\!\!\!:$ $D\h_{_{_{6d}}}\!\!\!\!\!=0$, 
{\bf SU(3) holonomy,} CS: $J_{ij}=i\h^\hc\g_{[i}\g_{j]}\h$, 
	~~~$DJ=0\then N_{ij}{}^k=0$ 
\\{\bf holomorphic 3-form} $\O\sim\h^\hc\g_{[i}\g_j\g_{k]}\h$,
{\bf Ricci flat K\"ahler} metric (DJ=0), vanishing $c_1$.
\centerline{	
	{\bf Mirror symmetry}: KS $\o_{\m\nb}\in H_{11}$ 
	~~$\leftrightarrow$~~ CS $\O_{\m\n\l}\d J^\l{}_{\bar\r}\in H_{21}$
}
\enddel

In the standard construction of anomaly free heterotic strings with gauge
group $E_6$ it turns out that charged particles and anti-particles
show up in conjunction with elements of the Dolbeault cohomology groups
$H^{11}$ and $H^{21}$, respectively. While $H^{11}$ parametrizes the
K\"ahler metrics, $H^{21}$ can be related to complex structure deformations
via contraction with the holomorphic 3-form, 
$\O_{\m\n\l}\d J^\l{}_{\bar\r}\in H^{21}$.
Since exchange of particles and anti-particles, as well as the 
corresponding sign of a $U(1)$ charge in the sigma model description of
the Calabi--Yau compactifications, are mere conventions, physicists 
came up with the idea of mirror symmetry \cite{lvw}, which was used by
Candelas et al. \cite{Candelas} to construct a mirror map between 
the K\"ahler and complex structure moduli spaces of a Calabi--Yau manifold
$X$ and its \hbox{mirror} dual $X^*$, whose topologies are related by
\BE							\label{TopMir}
	h_{11}(X)=h_{21}(X^*),\qquad	h_{21}(X)=h_{11}(X^*).
\EE
The power series expansions of this map could be interpreted as instanton
corrections in the quantum theory, and thus lead to a prediction of numbers of
rational curves \cite{Cox:vi,HKPTVZ}.

\subsection{Toric hypersurfaces}

The beauty of the toric construction of Calabi--Yau spaces is based
on the fact that it relates mirror symmetry to a combinatorial duality of
lattice polytopes, as was discovered by Batyrev \cite{Bat}. He showed that
the Calabi--Yau condition for a hypersurface, i.e. the vanishing of the 
first Chern class, requires as a necessary and sufficient condition that
the polytope $\D_D\subseteq M_\IR$ of the line bundle ${\cal O}(D)$ whose 
section defines the hypersurface is polar to the lattice polytope 
$\D^*=\D_D^\circ\subseteq N_\IR$ 
where $\D^*$ is the convex hull of the generators $v_j$ of the rays 
$\r_j\in\S^{(1)}$ of the fan of the ambient toric variety $X_\S$. 
A lattice polytope whose polar polytope (\ref{polar}) is again a lattice 
polytope is called 
{\it reflexive}. Batyrev also derived a combinatorial formula for the 
Hodge numbers
\BEA				\label{Bhodge}
&&     h_{11}(X_\D)=h_{2,1}(X_{\D^\circ})=
		l(\D^\circ)-1-\dim{\D}
\\&&	~~~~~~~-\!\!\!\!\sum_{\mathrm{codim}(\th^\circ)=1}l^*(\th^\circ)
		+\!\!\!\!
       \sum_{\mathrm{codim}(\th^\circ)=2}l^*(\th^\circ)l^*(\th)   \nonumber 
\EEA
where $\th$ and $\th^\circ$ is a dual pair of faces of $\D$ and $\D^\circ$, 
respectively. $l(\th)$ is
the number of lattice points of a face $\th$, and $l^*(\th)$ is the number of 
its interior lattice points. Mirror symmetry now amounts to the exchange 
of $\D$ and $\D^\circ$ and the formula for the Hodge data makes the 
topological duality (\ref{TopMir}) manifest.

The formula (\ref{Bhodge}) has a simple interpretation: The principal 
contributions to $h_{11}$ come from the toric divisors $D_j$ that correspond 
to lattice points in $\D^\circ$ different from the origin. 
There are $\dim(\D)$ linear relations among these divisors. The 
first sum corresponds to the subtraction of interior points of facets. 
The corresponding divisors of the ambient space do not intersect a generic 
Calabi--Yau hypersurface. Lastly, the bilinear terms in the second sum can 
be understood as multiplicities of toric divisors and their presence 
indicates that only a subspace of the K\"ahler moduli 
is accessible to toric methods.


\vbox to 283pt {\vspace{1pt}\begin{center}
		\newcount\hsum \newcount\hdif
\unitlength=0.44pt

\stepcounter{figure}						\makeatletter
\immediate\write\@auxout{\string\newlabel{4dRef}{\thefigure}} \makeatother
\end{center}\vspace*{4pt}
\begin{tabbing}
Fig. $\!\thefigure$: 
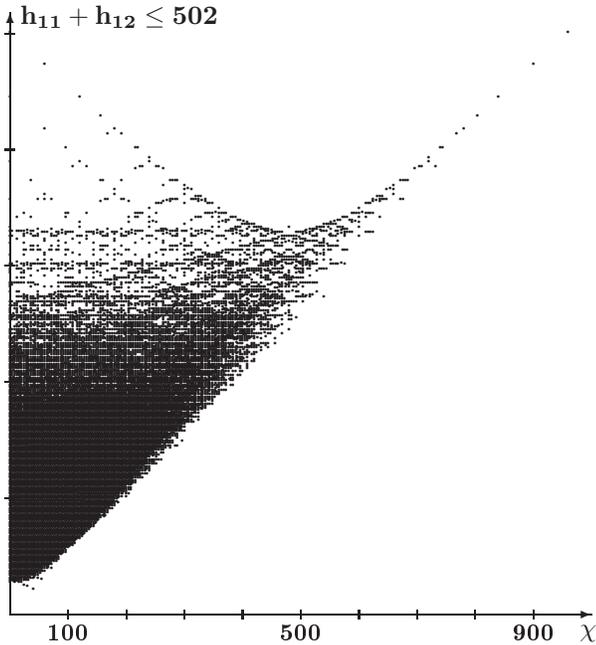
\=	Hypersurface spectra for $h_{11}\!\le\!h_{12}$. 
	\hbox{The maximal}\hfil
\\\>	$h_{11}+h_{12}$ comes from (251,251) and (491,11).
\end{tabbing}						    \vss } 

The enumeration of all reflexive polytopes has been achieved by Batyrev
for the two-dimensional case, as shown in figure \ref{TwoDimRP}, many years
ago. In dimensions 3 and 4, which are relevant for K3 surfaces and Calabi--Yau
3-folds, respectively, the enumeration required extensive use of computers and
was achieved in \cite{crp,c3d} and \cite{c4d,pwf}, respectively. The code
was later included into the software package PALP \cite{PALP}, which 
can be used for the reconstruction of the data as well as for many other
purposes like the analysis of fibrations and integral cohomology (see 
section \ref{Sec:FibTor}), as well as the construction of 
higher-dimensional examples. All results can be accessed on the web page
\cite{CYwien} and we just note that the numbers of reflexive polytopes in
3d and 4d are 4319 and 473\,800\,776, respectively. The resulting  
30108 different Hodge data of 3-folds amount to 15122 mirror pairs
as shown in figure \ref{4dRef}.

				 	\label{Sec:Hyper}
\vspace{-5pt}
\subsection{Complete intersections}  	\label{Sec:CICY}
\vspace{-2pt}
Soon after the hypersurface case Batyrev and Borisov discovered another 
beautiful combinatorial duality that corresponds to mirror symmetry of 
toric complete intersections \cite{BBnef}. In this generlization two 
polar pairs of
reflexive lattice polytopes are involved with the defining conditions
summarized in the following equations,
\BEA		\label{polyDN}
    \D=\D_1+\ldots+\D_r && \D^\circ=\langle\nabla_1,\ldots,\nabla_r\rangle
        _{\mao{conv}}\nonumber
\\[3pt]
        &\!\!\!\!\!\!\!\!\!(\D_l,\nabla_m)\ge-\delta_{lm} 
	\!\!\!\!\!\!\!\!\!\!\!\!\!\!\!\!\!&
\\[3pt]\nonumber
        \nabla^\circ=\langle\D_1,\ldots,\D_r\rangle
		_{\mao{conv}}\!\!\!\!\!\!\!\!\!
		 &&	\nabla=\nabla_1+\ldots+\nabla_r
\EEA
where $r$ is the codimension of the Calabi--Yau and the defining equations 
$f_i=0$ are sections of ${\cal O}(\D_i)$. The decomposition of 
the $M$-lattice polytope $\D\subset M_\IR$ into a Minkowski sum  
$\D=\D_1+\ldots+\D_r$ is now dual to a NEF 
(numerically effective) partition of the vertices of a different reflexive 
polytope $\nabla\subset N_\IR$ such that the convex hulls $\nabla_i$ of the
respective vertices and $0\in N$ only intersect at the origin
\cite{BBnef,Cox}.

The Hodge numbers $h_{pq}$  of the corresponding complete intersections 
have been computed and shown to obey (\ref{TopMir}) in \cite{BBsth}. 
They are summarized for arbitrary dimension $n-r$ of the Calabi--Yau 
in a generating polynomial $E(t,\bar t)$ as
\BEA                                          		\label{BBstrh}
    &&    \!\!\!\!\!\!\!\!E(t,\bar t)=\sum (-1)^{p+q}~h_{pq}~t^p\,\bar t^q
\\&&~~~~~=\sum_{I=[x,y]}
        \frac{(-)^{\rho_x}t^{\rho_y}}{(t\bar t)^r}S(C_x,\frac{\bar t}{t})
        S( C_y^\ast,t\bar t) B_I(t^{-1},\bar t) 	\nonumber
\EEA
in terms 
of the combinatorial data of the $n+r$ dimensional Gorenstein cone 
$\Gamma(\{\Delta_i\})$ spanned by vectors of the form $(e_i,v)$, where $e_i$ 
is a unit vector in $\IR^r$ and $v\in\Delta_i$. %
In this formula $x,y$ label faces $C_x$ of dimension $\rho_x$ of 
$\Gamma(\{\nabla_i\})$ and $C_x^\vee$ denotes the dual face of the dual cone 
of $\Gamma(\{\D_i\})$. The interval $I=[x,y]$ labels all cones that are 
faces of $C_y$ containing $C_x$. The {\it Batyrev--Borisov polynomials} 
$B_I(t,\bar t)$ encode certain 
combinatorial data of the face lattice \cite{BBsth}. The polynomials
$S(C_x,t)=(1-t)^{\rho_x}\sum_{n\ge0}t^nl_n(C_x)$ of degree $\rho_x-1$ are
related to the numbers $l_n(C_x)$ of lattice points at degree $n$ in $C_x$ 
and hence to the Ehrhart polynomial 
of the Gorenstein polytope generating $C_x$. 
(The Gorenstein polytope $\D_C$ consists of the degree 1 points of a
Gorenstein cone $C$. In the hypersurface case $r=1$ and $\D_C=\D=\D_1$.) 
\del
A Gorenstein cone $C$ is, by definition, generated by lattice points
$\rho_i$ with $\bra\rho_i,n_0\ket=1$ for some vector $n_0$ in the dual 
lattice. 
$n_0$ is unique if $C$ has maximal dimension and its duality pairing defines a
nonnegative grading of $C$. The points of degree 1 form the Gorenstein 
polytope, whose vertices generate $C$. 
\enddel

Without going into details let us emphasize that the formula
(\ref{BBstrh}) contains positive
and negative contributions whose interpretation in terms of individual 
contributions from toric divisors $D_j$ is, in contrast to the
hypersurface formula (\ref{Bhodge}), unfortunately 
unknown. In additional to efficiency problems in concrete calculations
(the formula is implemented in the nef-part of PALP \cite{PALP}, but
becomes quite slow for codimension $r>2$) this entails important theoretical 
problems, which we will comment on below.


\section{Fibrations and torsion in cohomology}		\label{Sec:FibTor}

Fibration structures play an important role in \hbox{string} theory, like
e.g. in heterotic--type II duality \cite{KachruV,KLM,AL,KKRS}, F-theory 
\cite{F,F12,Denef}, but also for the construction of vector bundles in
heterotic compactifications where they are often combined with
non-trivial fundamental groups \cite{Z3xZ3}. We now discuss
how these topological properties manifest themselves in combinatorial
properties of the polytopes that define toric Calabi--Yau varieties.

\subsection{Torsion in cohomology}

We begin with a discussion of the fundamental group, which is trivial
for every compact toric variety \cite{Dan} but may become non-trivial for
hypersurfaces and complete intersections. We first need to discuss smoothness
conditions and focus on the hypersurface case. If we 
consider the normal fan of a reflexive polytope $\D\subseteq M_\IR$ 
then $X_\S$ will generically have singularities which
have to be resolved if they have positive dimension while point-like
singularities can be avoided by a generic choice of the hypersurface equation.
The resolution can be performed by the choice of a convex (or {\it coherent})
triangulation of the fan $\S_\D$ \cite{BFS,GKZ,KKRS} whose rays should
consist of all rays over lattice points of $\D^\circ$ (this amounts
to a maximal star triangulation of $\D^\circ$; rays of lattice points in 
$N\setminus\D^\circ$ would contribute to $c_1$ and hence destroy the 
Calabi--Yau condition if the corresponding divisors intersect the 
hypersurface). For K3-surfaces such a triangulation already leads to a 
smooth toric ambient space because reflexivity implies that the facets are
at distance one and every maximal triangulation of a polygon consists of 
basic simplices. For CY 3-folds only the codimension-two cones of the
triangulation are basic while maximal-dimensional cones may contain 
pointlike singularities. This is still o.k. because pointlike singularities
can be avoided by a generic hypersurface. Toric 4-fold hypersurfaces, on
the other hand, may have terminal singularities that cannot be avoided,
so that many 5-dimensional reflexive polytopes cannot be used for the
construction of smooth CY hypersurfaces. For complete intersections the 
situation is analogous: 3-folds are generically smooth because the 
codimension $\dim \D-3$ of the Calabi--Yau is always larger than the
dimension  $\dim \D-4$ of the singular locus of $X_\S$. 

If we now consider a fixed polytope $\D\subset\IR^4$ without specification of
the lattice then reflexivity, i.e. integrality of the vertices of $\D$ and 
of $\D^\circ$, imply that $N$ is a sublattice of the dual of the lattice $M_V$
generated by the vertices of $\D$ and that $N$ contains the lattice $N_V$
generated by the vertices of $\D^\circ$ 
\BE
	N_V\subseteq N\subseteq M_V^\circ.		\label{Nchoice}
\EE
A refinement of the $N$ lattice amounts to a geometrical quotient by a 
group action $G$, which we call {\it toric} because it acts diagonally 
on the homogeneous coordinates.
Such a refinement always 
entails additional quotient singularities in the ambient space and no 
contributions to its fundamental group \cite{Dan}. If, however, a 
Calabi--Yau does not intersect the singular locus of that quotient then the 
group acts freely on that variety and contributes to $\pi_1$. 
This is the case if the refinement of the lattice does not lead to 
additional lattice points of $\D^\circ$ (more precisely, lattice points
of codimension one can be ignored because, according to eq. (\ref{Bhodge}), 
the corresponding divisors do not intersect the hypersurface).

For a given pair of reflexive polytopes there is only a finite number of
lattices $N$ that obey (\ref{Nchoice}) and hence only a finite number of
possible toric free quotients. In \cite{BK} we have shown that the 
fundamental group of a toric CY hypersurface is isomorphic to the lattice
quotient of the $N$-lattice divided by its sublattice $N^{(3)}$ generated 
by the lattice points on 3-faces of $\S$. 
All fundamental groups of toric hypersurfaces thus come from toric quotients
and are abelian, so that $\p_1$ is isomorphic to the torsion in $H^2$.
We also found a combinatorial formula for the Brauer group $B$, which
is the torsion the third cohomology $H^3$, in terms of the sublattice 
$N^{(2)}$ generated by lattice points on 2-faces of $\S$. Here, however,
$B\ex B$ must be a subgroup of $N/N^{(2)}$.

Various dualities imply that the complete torsion in the cohomology group
is determined in terms of $\p_1$ and $B$ and we conjectured, based on some 
$K$-theory arguments, that these groups are exchanged with the duals of 
the other under mirror duality \cite{BK}. This conjecture could be 
verified for all toric hypersurfaces by explicit calculation. 
Two well-known examples are the free $\IZ_5$ quotient of the quintic and the
free $\IZ_3$ quotient of the CY hypersurface in $\IP^2\ex\IP^2$.
For the
complete \hbox{list} of 473\,800\,776 reflexive polytopes
one finds 14 more examples of toric free quotients \cite{PALP}: The 
elliptically fibered $\IZ_3$ 
quotient of the degree $9$ surface in $\IP^4_{11133}$, whose group action 
on the homogeneous coordinates is given by the phases $(1,2,1,2,0)/3$,
and 13 elliptic K3 fibrations where the lattice quotient has index 2.
The 16 non-trivial Brauer groups showed up, as expected, 
exactly for the 16 polytopes that are polar to the ones that lead to a
non-trivial fundamental group. For complete intersections it is possible
to have both a non-trivial fundamental group and a non-trivial Brauer
group at the same time and our conjecture was verified
for a codimension-2 Calabi--Yau for which both groups are 
$\IZ_3\ex\IZ_3$ \cite{Braun}.

\del
Torsion \cite{BK,Gross,Braun,Z3xZ3,AMcring}

Fibrations \cite{oguiso}

\subsection*{Triangulation}		\cite{BFS,GKZ,KKRS}

CICYs \cite{BBnef,BBsth,wpci,KKRS}

\enddel

\subsection{Fibrations}

For general K3 surfaces and Calabi--Yau 3-folds
there exists a criterion by Oguiso for the existence of elliptic and K3
fibrations in terms of intersection numbers \cite{oguiso,AL}. In the toric
context the data of a given reflexive polytope $\D^\circ\subseteq N_\IR$ has
to be supplemented by a triangulation of the fan, as discussed above, 
and the fibration properties like intersections depend on the 
chosen triangulation. 

Computation of all intersection numbers for all triangulations is 
computationally quite expensive, but 
for toric Calabi--Yau spaces there is, fortunately, a more direct way to 
search for fibrations that manifest themselves in the geometry of the 
polytope 
and to single out the appropriate 
triangulations~\cite{AKMS,fft,Hu:2000,pwf,Rohsie}. 
These fibrations descend from toric morphisms of the ambient
space \cite{Ewald,Ful}, which correspond to a map $\phi:\S\to\S_b$ of 
fans in $N$ and $N_b$, respectively,
where $\phi:N\to N_b$ is a lattice homomorphism such that for each cone 
$\s\in\S$ there is a cone $\s_b\in\S_b$ that contains the image of $\s$. 
The lattice $N_f$ for the fiber is the kernel of $\phi$ in $N$.

\vbox to 130pt	{\vss\begin{center}
\unitlength=2.2pt
\BP(60,25)				\def\Y{21}	\def\H{18}\def\X{6}
\put(30,22){\drawline(0,-\H)(-\Y,0)\drawline(0,-\H)(\Y,0)
        \drawline(0,-\H)(-10,-\X)\drawline(0,-\H)(10,-\X)
        \dashline2(0,-\H)(10,\X)\dashline2(0,-\H)(-10,\X)
        \drawline(0,\H)(-\Y,0)\drawline(0,\H)(\Y,0)
        \drawline(0,\H)(-10,-\X)\drawline(0,\H)(10,-\X)
        \dashline2(0,\H)(10,\X)\dashline2(0,\H)(-10,\X)
        }
\put(30,22){\lila\linethickness{2pt}
        \drawline(-\Y,0)(-10,\X)\drawline(10,\X)(-10,\X)\drawline(10,\X)(\Y,0)
    \drawline(-\Y,0)(-10,-\X)\drawline(10,-\X)(-10,-\X)\drawline(10,-\X)(\Y,0)
\black  }
\EP
\\					\stepcounter{figure}	\makeatletter
\immediate\write\@auxout{\string\newlabel{Fibration}{\thefigure}}\makeatother
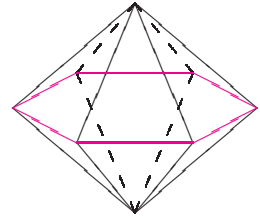
Fig. \thefigure: CY fibration from reflexive sections
		of $\D^\circ\subseteq N_\IR$.
\end{center}\vspace{-5pt}}

If we are interested in fibrations whose fibers are Calabi--Yau varieties
of lower dimension then the restriction of the defining equations to the
fan $\S_f$ in $N_f$ needs to satisfy Batyrev's criterion. We hence
require that the intersection $\D^\circ_f=\D^\circ\cap N_f$ is reflexive,
like the intersection with the horizontal plane in figure \ref{Fibration}.
The search for toric fibrations hence amounts to a search for 
reflexive sections of $\D^\circ$ with appropriate dimension
(or, equivalently, for reflexive projections in the $M$-lattice, which was
used in a search for K3 fibrations in \cite{AKMS}). In order to guarantee the 
existence of the projection we choose a triangulation of $\D^\circ_f$ and then
extend it to a triangulation of $\D^\circ$ (this may not always be possible
if the codimension is larger that 1, as was pointed out and analyzed by 
Rohsiepe \cite{Rohsie}).
For each such choice we can interpret the homogeneous coordinates that 
correspond to rays in $\D_f^\circ$ 
as coordinates of the fiber and the others as 
parameters of the equations and hence as moduli of the fiber space.

For hypersurfaces the geometry of the resulting fibration has been worked 
out in detail in~\cite{Hu:2000}. Even in the case of complete intersetions 
reflexivity of the fiber 
polytope $\D_f^\circ$ ensures that the fiber also is a complete intersection
Calabi--Yau because a 
nef partition of $\D^\circ$ automatically
induces a nef partition of $\D_f^\circ$ \cite{KKRS}.
The codimension $r_f$ of the fiber generically 
coincides with the codimension $r$ of the fibered space but for 
$r>1$ it may happen that $\S_f^1$ does not 
intersect one (or more) of the $\D_i$ of the nef partition, in which case 
the codimension decreases \cite{KKRS}. In \cite{KKRS} we performed
an extensive search for K3-fibrations in complete intersections which, 
due to modular properties, could be used for all-genus calculations of
topological string amplitudes. In this search we also encountered an
example where the fibration does not extend to a morphism of the ambient
spaces because some exceptional points do not intersect the Calabi--Yau.
Even in that example, however, the K3 fiber is realized by a fan on
a sublattice.

\section{Work in progress and open problems}		\label{Sec:Conc}

For toric Calabi--Yau hypersurfaces in 3 dimensions the enumeration and
the computation of the integral cohomology has been completed, but for 
the case of complete intersections only the surface habe been scratched
\cite{wpci,KKRS}.
While the number of reflexive polytopes in 5 dimensions, which would
be relevant for 4-folds as used in F-theory, is simply to large
(maybe something like $10^{18}$) there is some hope that a classification
of complete intersection 3-folds may be feasable, at least for small 
codimensions, via an enumeration of reflexive Gorenstein cones 
\cite{BBgen,srp}. 
On the theoretical side it would be important to find a better
formula for the Hodge data that allows a direct interpretation of the
Picard number in terms of toric divisors (for codimension $r>1$ even
divisors that correspond to vertices of $\nabla$ may not intersect the
Calabi--Yau \cite{KKRS}). A related issue is the search for a
combinatorial formula for torsion in cohomology, which would also be very 
useful for model building.

\vbox to 172pt	{\vss	\unitlength=2.7pt	\def\Msum{46}
\begin{center}	\def\putlin#1,#2,#3,#4,#5){\put#1,#2){\line(#3,#4){#5}}}
		\def\putvec#1,#2,#3,#4,#5){\put#1,#2){\vector(#3,#4){#5}}}
	\def\Xlab#1 {\put(#1,#1){\drawline(-.5,.5)(.5,-.5)}}
	\def\Ylab#1 {\put(-#1,#1){\drawline(.5,.5)(-.5,-.5)}}

\BP(80,25)(-40,2)	
	\putlin(0,0,1,1,44.5)	
	\putlin(0,0,-1,1,44.5)	
	\Xlab10	\Xlab20	\Xlab30	\Xlab40	\Ylab10	\Ylab20	\Ylab30	\Ylab40	
        \def\h#1.#2.{     \hsum=#1 \advance\hsum by #2 
                \hdif=-#2 \advance\hdif by #1
	\ifnum \Msum < \hsum \else	\put(\hdif,\hsum){\blue\circle{1.4}}
					\put(-\hdif,\hsum){\blue\circle{1.4}}
	\fi}
\h1.25.\h1.28.\h1.29.\h1.30.\h1.31.\h1.32.\h1.33.\h1.34.\h1.35.\h1.36.\h1.37.
	\h1.38.\h1.39.\h1.40.\h1.41.\h1.45.\h1.47.\h1.51.\h1.53.\h1.55.\h1.59.

\h2.26.\h2.28.\h2.29.\h2.30.\h2.31.\h2.32.\h2.33.\h2.34.\h2.35.\h2.36.\h2.37.
	\h2.38.\h2.39.\h2.40.\h2.41.\h2.42.\h2.43.\h2.44.\h2.45.\h2.46.\h2.47.
\h3.25.\h3.27.\h3.28.\h3.29.\h3.30.\h3.31.\h3.32.\h3.33.\h3.34.\h3.35.\h3.36.
	\h3.37.\h3.38.\h3.39.\h3.40.\h3.41.\h3.42.\h3.43.\h3.44.\h3.45.\h3.46.
\h4.24.\h4.28.\h4.30.\h4.31.\h4.32.\h4.33.\h4.34.\h4.35.\h4.36.\h4.37.\h4.38.
	\h4.39.\h4.40.\h4.41.\h4.42.\h4.43.\h4.44.\h4.45.\h4.46.\h4.47.\h4.48.
\h5.27.\h5.29.\h5.30.\h5.31.\h5.32.\h5.33.\h5.34.\h5.35.\h5.36.\h5.37.\h5.38.
	\h5.39.\h5.40.\h5.41.\h5.42.\h5.43.\h5.44.\h5.45.\h5.46.\h5.47.
\h6.28.\h6.30.\h6.31.\h6.32.\h6.34.\h6.35.\h6.36.\h6.37.\h6.38.\h6.39.\h6.40.
	\h6.41.\h6.42.\h6.43.\h6.44.\h6.45.\h6.46.\h6.47.
\h7.27.\h7.29.\h7.30.\h7.31.\h7.33.\h7.34.\h7.35.\h7.37.\h7.38.\h7.39.\h7.40.
	\h7.41.\h7.43.\h7.45.\h7.47.\h7.49.\h7.50.\h7.51.
\h8.30.\h8.32.\h8.33.\h8.34.\h8.36.\h8.38.\h8.40.\h8.42.\h8.44.\h8.52.
\h9.31.\h9.33.\h9.37.	\h10.26.\h10.30.\h10.34.\h10.36.
\h11.27.	\h12.28.	\h15.23.
        \def\h#1.#2.{     \hsum=#1 \advance\hsum by #2 
                \hdif=-#2 \advance\hdif by #1
\ifnum \Msum < \hsum \else	\put(\hdif,\hsum){\black\circle*{1}}
				\put(-\hdif,\hsum){\black\circle*{1}}	\fi}
\h14.14.\h15.15.\h16.12.\h16.13.\h16.14.\h16.15.\h16.16.\h17.12.
\h17.13.\h17.14.\h17.15.\h17.16.\h17.17.\h18.10.\h18.11.\h18.12.
\h18.13.\h18.14.\h18.15.\h18.16.\h18.17.\h18.18.\h19.7.\h19.9.
\h19.11.\h19.13.\h19.14.\h19.15.\h19.16.\h19.17.\h19.18.\h19.19.
\h20.5.\h20.10.\h20.11.\h20.12.\h20.13.\h20.14.\h20.15.\h20.16.
\h20.17.\h20.18.\h20.19.\h20.20.\h21.1.\h21.9.\h21.11.\h21.12.
\h21.13.\h21.14.\h21.15.\h21.16.\h21.17.\h21.18.\h21.19.\h21.20.
\h21.21.\h22.8.\h22.9.\h22.10.\h22.11.\h22.12.\h22.13.\h22.14.
\h22.15.\h22.16.\h22.17.\h22.18.\h22.19.\h22.20.\h22.21.\h22.22.
\h23.7.\h23.9.\h23.10.\h23.11.\h23.12.\h23.13.\h23.14.\h23.15.
\h23.16.\h23.17.\h23.18.\h23.19.\h23.20.\h23.21.\h23.22.\h23.23.
\h24.8.\h24.9.\h24.10.\h24.11.\h24.12.\h24.13.\h24.14.\h24.15.
\h24.16.\h24.17.\h24.18.\h24.19.\h24.20.\h24.21.\h24.22.\h24.23.
\h24.24.\h25.7.\h25.8.\h25.9.\h25.10.\h25.11.\h25.12.\h25.13.
\h25.14.\h25.15.\h25.16.\h25.17.\h25.18.\h25.19.\h25.20.\h25.21.
\h25.22.\h25.23.\h25.24.\h25.25.\h26.6.\h26.8.\h26.9.\h26.10.
\h26.11.\h26.12.\h26.13.\h26.14.\h26.15.\h26.16.\h26.17.\h26.18.
\h26.19.\h26.20.\h26.21.\h26.22.\h26.23.\h26.24.\h26.25.\h26.26.
\h27.6.\h27.7.\h27.8.\h27.9.\h27.10.\h27.11.\h27.12.\h27.13.
\h27.14.\h27.15.\h27.16.\h27.17.\h27.18.\h27.19.\h27.20.\h27.21.
\h27.22.\h27.23.\h27.24.\h27.25.\h27.26.\h27.27.\h28.4.\h28.6.
\h28.7.\h28.8.\h28.9.\h28.10.\h28.11.\h28.12.\h28.13.\h28.14.
\h28.15.\h28.16.\h28.17.\h28.18.\h28.19.\h28.20.\h28.21.\h28.22.
\h28.23.\h28.24.\h28.25.\h28.26.\h28.27.\h28.28.\h29.2.\h29.5.
\h29.7.\h29.8.\h29.9.\h29.10.\h29.11.\h29.12.\h29.13.\h29.14.
\h29.15.\h29.16.\h29.17.\h29.18.\h29.19.\h29.20.\h29.21.\h29.22.
\h29.23.\h29.24.\h29.25.\h29.26.\h29.27.\h29.28.\h29.29.\h30.6.
\h30.7.\h30.8.\h30.9.\h30.10.\h30.11.\h30.12.\h30.13.\h30.14.
\h30.15.\h30.16.\h30.17.\h30.18.\h30.19.\h30.20.\h30.21.\h30.22.
\h30.23.\h30.24.\h30.25.\h30.26.\h30.27.\h30.28.\h30.29.\h30.30.
\h31.5.\h31.7.\h31.8.\h31.9.\h31.10.\h31.11.\h31.12.\h31.13.
\h31.14.\h31.15.\h31.16.\h31.17.\h31.18.\h31.19.\h31.20.\h31.21.
\h31.22.\h31.23.\h31.24.\h31.25.\h31.26.\h31.27.\h31.28.\h31.29.
\h31.30.\h31.31.\h32.6.\h32.7.\h32.8.\h32.9.\h32.10.\h32.11.
\h32.12.\h32.13.\h32.14.\h32.15.\h32.16.\h32.17.\h32.18.\h32.19.
\h32.20.\h32.21.\h32.22.\h32.23.\h32.24.\h32.25.\h32.26.\h32.27.
\h32.28.\h32.29.\h32.30.\h32.31.\h32.32.\h33.5.\h33.6.\h33.7.
\h33.8.\h33.9.\h33.10.\h33.11.\h33.12.\h33.13.\h33.14.\h33.15.
\h33.16.\h33.17.\h33.18.\h33.19.\h33.20.\h33.21.\h33.22.\h33.23.
\h33.24.\h33.25.\h33.26.\h33.27.\h33.28.\h33.29.\h33.30.\h33.31.
\h33.32.\h33.33.\h34.4.\h34.6.\h34.7.\h34.8.\h34.9.\h34.10.
\h34.11.\h34.12.\h34.13.\h34.14.\h34.15.\h34.16.\h34.17.\h34.18.
\h34.19.\h34.20.\h34.21.\h34.22.\h34.23.\h34.24.\h34.25.\h34.26.
\h34.27.\h34.28.\h34.29.\h34.30.\h34.31.\h34.32.\h34.33.\h34.34.
\h35.5.\h35.7.\h35.8.\h35.9.\h35.10.\h35.11.\h35.12.\h35.13.
\h35.14.\h35.15.\h35.16.\h35.17.\h35.18.\h35.19.\h35.20.\h35.21.
\h35.22.\h35.23.\h35.24.\h35.25.\h35.26.\h35.27.\h35.28.\h35.29.
\h35.30.\h35.31.\h35.32.\h35.33.\h35.34.\h35.35.\h36.4.\h36.6.
\h36.7.\h36.8.\h36.9.\h36.10.\h36.11.\h36.12.\h36.13.\h36.14.
\h36.15.\h36.16.\h36.17.\h36.18.\h36.19.\h36.20.\h36.21.\h36.22.
\h36.23.\h36.24.\h36.25.\h36.26.\h36.27.\h36.28.\h36.29.\h36.30.
\h36.31.\h36.32.\h36.33.\h36.34.\h36.35.\h36.36.\h37.4.\h37.5.
\h37.6.\h37.7.\h37.8.\h37.9.\h37.10.\h37.11.\h37.12.\h37.13.
\h37.14.\h37.15.\h37.16.\h37.17.\h37.18.\h37.19.\h37.20.\h37.21.
\h37.22.\h37.23.\h37.24.\h37.25.\h37.26.\h37.27.\h37.28.\h37.29.
\h37.30.\h37.31.\h37.32.\h37.33.\h37.34.\h37.35.\h37.36.\h37.37.
\h38.2.\h38.5.\h38.6.\h38.7.\h38.8.\h38.9.\h38.10.\h38.11.
\h38.12.\h38.13.\h38.14.\h38.15.\h38.16.\h38.17.\h38.18.\h38.19.
\h38.20.\h38.21.\h38.22.\h38.23.\h38.24.\h38.25.\h38.26.\h38.27.
\h38.28.\h38.29.\h38.30.\h38.31.\h38.32.\h38.33.\h38.34.\h38.35.
\h38.36.\h38.37.\h38.38.\h39.5.\h39.6.\h39.7.\h39.8.\h39.9.
\h39.10.\h39.11.\h39.12.\h39.13.\h39.14.\h39.15.\h39.16.\h39.17.
\h39.18.\h39.19.\h39.20.\h39.21.\h39.22.\h39.23.\h39.24.\h39.25.
\h39.26.\h39.27.\h39.28.\h39.29.\h39.30.\h39.31.\h39.32.\h39.33.
\h39.34.\h39.35.\h39.36.\h39.37.\h39.38.\h39.39.\h40.4.\h40.5.
\h40.6.\h40.7.\h40.8.\h40.9.\h40.10.\h40.11.\h40.12.\h40.13.
\h40.14.\h40.15.\h40.16.\h40.17.\h40.18.\h40.19.\h40.20.\h40.21.
\h40.22.\h40.23.\h40.24.\h40.25.\h40.26.\h40.27.\h40.28.\h40.29.
\h40.30.\h40.31.\h40.32.\h40.33.\h40.34.\h40.35.\h40.36.\h40.37.
\h40.38.\h40.39.\h40.40.\h41.5.\h41.6.\h41.7.\h41.8.\h41.9.
\h41.10.\h41.11.\h41.12.\h41.13.\h41.14.\h41.15.\h41.16.\h41.17.
\h41.18.\h41.19.\h41.20.\h41.21.\h41.22.\h41.23.\h41.24.\h41.25.
\h41.26.\h41.27.\h41.28.\h41.29.\h41.30.\h41.31.\h41.32.\h41.33.
\h41.34.\h41.35.\h41.36.\h41.37.\h41.38.\h41.39.\h41.40.\h41.41.
\h42.4.\h42.5.\h42.6.\h42.7.\h42.8.\h42.9.\h42.10.\h42.11.
\h42.12.\h42.13.\h42.14.\h42.15.\h42.16.\h42.17.\h42.18.\h42.19.
\h42.20.\h42.21.\h42.22.\h42.23.\h42.24.\h42.25.\h42.26.\h42.27.
\h42.28.\h42.29.\h42.30.\h42.31.\h42.32.\h42.33.\h42.34.\h42.35.
\h42.36.\h42.37.\h42.38.\h42.39.\h42.40.\h42.41.\h42.42.\h43.3.
\h43.5.\h43.6.\h43.7.\h43.8.\h43.9.\h43.10.\h43.11.\h43.12.
\h43.13.\h43.14.\h43.15.\h43.16.\h43.17.\h43.18.\h43.19.\h43.20.
\h43.21.\h43.22.\h43.23.\h43.24.\h43.25.\h43.26.\h43.27.\h43.28.
\h43.29.\h43.30.\h43.31.\h43.32.\h43.33.\h43.34.\h43.35.\h43.36.
\h43.37.\h43.38.\h43.39.\h43.40.\h43.41.\h43.42.\h43.43.\h44.4.
\h44.5.\h44.6.\h44.7.\h44.8.\h44.9.\h44.10.\h44.11.\h44.12.
\h44.13.\h44.14.\h44.15.\h44.16.\h44.17.\h44.18.\h44.19.\h44.20.
\h44.21.\h44.22.\h44.23.\h44.24.\h44.25.\h44.26.\h44.27.\h44.28.
\h44.29.\h44.30.\h44.31.\h44.32.\h44.33.\h44.34.\h44.35.\h44.36.
\h44.37.\h44.38.\h44.39.\h44.40.\h44.41.\h44.42.\h44.43.\h44.44.
\h45.3.\h45.5.\h45.6.\h45.7.\h45.8.\h45.9.\h45.10.\h45.11.
\h45.12.\h45.13.\h45.14.\h45.15.\h45.16.\h45.17.\h45.18.\h45.19.
\h45.20.\h45.21.\h45.22.\h45.23.\h45.24.\h45.25.\h45.26.\h45.27.
\h45.28.\h45.29.\h45.30.\h45.31.\h45.32.\h45.33.\h45.34.\h45.35.
\h45.36.\h45.37.\h45.38.\h45.39.\h45.40.\h45.41.\h45.42.\h45.43.
\h45.44.\h45.45.\h46.4.\h46.5.\h46.6.\h46.7.\h46.8.\h46.9.
\h46.10.\h46.11.\h46.12.\h46.13.\h46.14.\h46.15.\h46.16.\h46.17.
\h46.18.\h46.19.\h46.20.\h46.21.\h46.22.\h46.23.\h46.24.\h46.25.
\h46.26.\h46.27.\h46.28.\h46.29.\h46.30.\h46.31.\h46.32.\h46.33.
\h46.34.\h46.35.\h46.36.\h46.37.\h46.38.\h46.39.\h46.40.\h46.41.
\h46.42.\h46.43.\h46.44.\h46.45.\h46.46.\h47.5.\h47.6.\h47.7.
\h47.8.\h47.9.\h47.10.\h47.11.\h47.12.\h47.13.\h47.14.\h47.15.
\h47.16.\h47.17.\h47.18.\h47.19.\h47.20.\h47.21.\h47.22.\h47.23.
\h47.24.\h47.25.\h47.26.\h47.27.\h47.28.\h47.29.\h47.30.\h47.31.
\h47.32.\h47.33.\h47.34.\h47.35.\h47.36.\h47.37.\h47.38.\h47.39.
\h47.40.\h47.41.\h47.42.\h47.43.\h47.44.\h47.45.\h47.46.\h47.47.
\h48.4.\h48.5.\h48.6.\h48.7.\h48.8.\h48.9.\h48.10.\h48.11.
\h48.12.\h48.13.\h48.14.\h48.15.\h48.16.\h48.17.\h48.18.\h48.19.
\h48.20.\h48.21.\h48.22.\h48.23.\h48.24.\h48.25.\h48.26.\h48.27.
\h48.28.\h48.29.\h48.30.\h48.31.\h48.32.\h48.33.\h48.34.\h48.35.
\h48.36.\h48.37.\h48.38.\h48.39.\h48.40.\h48.41.\h48.42.\h48.43.
\h48.44.\h48.45.\h48.46.\h48.47.\h48.48.\h49.4.\h49.5.\h49.6.
\h49.7.\h49.8.\h49.9.\h49.10.\h49.11.\h49.12.\h49.13.\h49.14.
\h49.15.\h49.16.\h49.17.\h49.18.\h49.19.\h49.20.\h49.21.\h49.22.
\h49.23.\h49.24.\h49.25.\h49.26.\h49.27.\h49.28.\h49.29.\h49.30.
\h49.31.\h49.32.\h49.33.\h49.34.\h49.35.\h49.36.\h49.37.\h49.38.
\h49.39.\h49.40.\h49.41.\h49.42.\h49.43.\h49.44.\h49.45.\h49.46.
\h49.47.\h49.48.\h49.49.\h50.4.\h50.5.\h50.6.\h50.7.\h50.8.
\EP\end{center}
					\stepcounter{figure}	\makeatletter
\immediate\write\@auxout{\string\newlabel{Conifolds}{\thefigure}}\makeatother
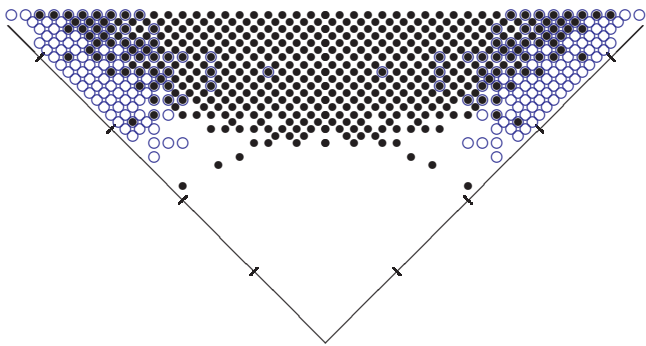
\begin{tabbing}Fig. \thefigure: \=Deformed conifold Hodge data (circles)
	and toric 
\\	\>CY hypersurfaces (dots) with $h_{11}+h_{12}\le\Msum$.
\end{tabbing}\vspace{-5pt}}

Inspite of the fact that the toric construction yield by far the largest
class of known Calabi--Yau spaces, it is unclear how generic these spaces
are and it is not even known whether the total number of topological types
is finite \cite{EllCY}. A first step beyond the toric realm along the lines 
of \cite{duco} has been taken recently when we studied conifold 
transition to non-toric Calabi--Yau spaces \cite{Coni}.
As shown in figure \ref{Conifolds} this construction, 
which still uses toric tools,
yields a surprisingly rich class of new Calabi--Yau spaces with small
Picard number $h_{11}$. The realm with small $h_{11}+h_{21}$, on the other
hand, seems to be poppulated by varieties with non-trivial fundamental group
\cite{Gross}. Systematic studies of free quotients, however, so far have 
only been performed in special cases.


\def\bibadd{
}

\del
The condition that $\D^*\cap N$ and $\D^*\cap N'$ coincide is sufficient for 
a free quotient of a CICY with dimension up to 3
because the singularities of a maximal crepant resolution are at 
codimension 4 and can be avoided by a generic choice of the defining 
equations. It is, however, not necessary:  Divisors corresponding 
to interior points of facets of $\D^*$ do not intersect the CY and hence 
do not kill the fundamental group if they are generated by a refinement
of the $N$ lattice \cite{BK}.

{\footnotesize\begin{verbatim}
Why Calabi-Yau?
        SUSY: SU(3) holonomy $\then$ complex / KM-dhler / Ricci-flat (1st Chern)
        mirror symmetry, special KM-dhler
Why toric geometry?
        compact / scaling (patching) ... projective weighted toric
        physics interpretation: complex / symplectic geometry
Constructions: Batyrev / Borisov ... combinatorics
        physics::topological string, fundamental groups ...,
        fibrations, K-theory::Brauer group,
        PALP ...

Strings, branes and dualities
        substructure::Kaluza-Klein
                CFT:: Wick rotation, contraction, ghosts ... Casimir::c/12
                path integral//insertions in bulk/boundary
        RNS / sigma model on CY: N=2 SCFT
        special geometry ... (exists prepotential, flat coordinates 
                                instanton sums and mirror symmetry)
        superstrings::Berkovits::SUGRA
        compactifications and dualities T ... D-branes = SUGRA instantons ... 
        RR-flux and generalized complex geometry ...
\end{verbatim}}
\enddel

{\it Acknowledgements.} I would like to thank Stephan Moskaliuk for the 
invitation to the Bogolyubov Institute for Theoretical Physics in Kiev, 
where part of these notes were presented at the Kummer Symposium 
``Fundamental Problems in Modern Quantum Theories and Experiments III''. 
It is also a pleasure to  acknowledge many vital and enjoyable 
discussions with my collaborators Victor Batyrev, Albrecht Klemm, 
Erwin Riegler, Emanuel Scheidegger and Harald Skarke.
This work was supported in part by the Austrian Research Funds FWF under
grant number P18679-N16.

\end{multicols}

\end{document}